\documentclass[letter,12pt]{article}

\RequirePackage{amsthm,amsmath,amsfonts,amssymb,bm, xcolor,graphicx,mathrsfs}

\RequirePackage{multirow,epstopdf}
\usepackage{afterpage}
\usepackage{natbib}
\usepackage{verbatim}
\usepackage{longtable}
\usepackage{threeparttable}
\usepackage{color}
\usepackage{amsmath}
\usepackage{booktabs}
\numberwithin{equation}{section}

\newtheorem{lemma}{Lemma}
\newtheorem{theorem}{Theorem}
\newtheorem{remark}{Remark}
\newtheorem{corollary}{Corollary}
\renewcommand{\theequation}{\thesection.\arabic{equation}}

\renewcommand{\hat}{\widehat}

\def\singlespace{\def\baselinestretch{1}\@normalsize}

\makeatother

\newcommand{\bY}{\mathbf{Y}}

\newcommand{\bX}{\mathbf{X}}
\newcommand{\bZ}{\mathbf{Z}}

\newcommand{\bV}{\mathbf{V}}

\newcommand{\bM}{\mathbf{M}}
\newcommand{\bR}{\mathbf{R}}
\newcommand{\bN}{\mathbf{N}}
\newcommand{\bbeta}{\bm{\beta}}

\newcommand{\btheta}{\bm{\theta}}

\newcommand{\bvarth}{\bm{\vartheta}}
\newcommand{\bdelta}{\bm{\delta}}

\def\newpage{\vfill\eject}

\newdimen\biblioindent    \biblioindent=30pt

 at 7truept
 
\def\sgn{\mbox{sgn}}

\def\bH{\bm{H}}

\def\bV{\bm{V}}

\def\beq{\begin{equation}}
\def\eeq{\end{equation}}
\def\beqn{\begin{eqnarray}}
\def\eeqn{\end{eqnarray}}
\def\beqnn{\begin{eqnarray*}}
\def\eeqnn{\end{eqnarray*}}

\def\bal{\bm{\alpha}}

\def\A{{\mathcal A}}
\usepackage{color}

\allowdisplaybreaks

\begin{document}
\def\spacingset#1{\renewcommand{\baselinestretch}%
{#1}\small\normalsize} \spacingset{1}


\title{\bf Estimation and Inference in Ultrahigh Dimensional Partially Linear Single-Index
Models\thanks{All authors make equally contribution to this work and are listed in the alphabetic order.}}

\author{Shijie Cui$^a$, Xu Guo$^b$ and Zhe Zhang$^c$\\
 $^a$ Department of Statistics, Pennsylvania State University\\
University Park, PA, USA\\
$^b $School of Statistics, Beijing Normal University\\ Beijing, 100875, P. R. China \\
$^c $University of North Carolina at Chapel Hill\\ Chapel Hill, NC, 27599, USA}
\date{}
\maketitle

\begin{abstract}
This paper is concerned with estimation and inference for ultrahigh dimensional partially
linear single-index models. The presence of high dimensional nuisance parameter
and nuisance unknown function makes the estimation and inference problem very
challenging. In this paper, we first
propose a profile partial penalized least squares estimator and establish the sparsity,
consistency and asymptotic representation of the proposed estimator in ultrahigh dimensional setting. We then propose an $F$-type test statistic for parameters of
primary interest and show that the limiting null distribution of the test
statistic is $\chi^2$ distribution, and the test statistic can detect local
alternatives, which converge to the null hypothesis at the root-$n$ rate. We
further propose a new test for the specification testing problem of the
nonparametric function. The test statistic is shown to be asymptotically
normal. Simulation studies are conducted to examine the finite sample
performance of the proposed estimators and tests. A real data example is used to illustrate
the proposed procedures.
\end{abstract}
\noindent%
\vfill

\newpage
\spacingset{1.45} 

\pagestyle{plain}

\section{Introduction}
Thanks to advances in computing technologies, high dimensional
modeling has been increasingly common in economics and finance,
including microeconomics, macroeconomics, marketing, and portfolio
selection as well as other areas such as medical studies and
health studies.  See \cite{Fan:Li:Zhang:Zou:2020} for an overview. 
In high dimensional analysis, statistical inference has attracted
considerable attention. Recently there are many developments for
testing low dimensional parametric components in high dimensional
models, see for instance 
\cite{barber2015controlling}, \cite{lan2016testing}, \cite{ning2017general}, \cite{candes2016panning}
\cite{shi2019linear}, \cite{liu2020model} and references therein. However, these works
focused on high dimensional parametric modeling. It is unknown
whether their works apply to more general settings, e.g.
semiparametric setting.

Compared with parametric regression modeling, semiparametric
models relax restrictive assumptions on parametric
models and are flexible enough to capture the relationship between
the response and the covariates. The partially linear single-index
model (PLSIM) is one of the popular semiparametric regression
models and is widely used in economics. See, for example, Example
1.1.6 in \citet{hardle2012partially} for an interesting
application of PLSIM.
%

Let $Y$ be the response, $\bX$ and $\bZ$ be the $p$- and $q$-dimensional
covariates, respectively. In this paper, $q$ is fixed while $p$ is allowed to be exponential order of the sample size.
Thus we consider the ultrahigh dimensional setting.
The PLSIM is
\begin{eqnarray}\label{eqn1.1}
Y=\eta(\bal^\top \bX)+\bbeta^\top  \bZ+\epsilon,
\end{eqnarray}
where $\bal$ and $\bbeta$ are unknown parameters, $\eta(\cdot)$ is an unknown
smooth function, $E(\epsilon|\bX,\bZ)=0$, and $E(\epsilon^2|\bX,\bZ)=\sigma^2$. For
model identification, assume that $\|\bal\|_2=1$ and its first element is
positive, where $\|\cdot\|_2$ is the $L_2$ norm. Model (\ref{eqn1.1}) is quite
general. When $p=1$, it reduces to the partially linear model
\citep{speckman1988kernel} while it turns to be the single-index model \citep{ichimura1993semiparametric} when $\bbeta=0$.

To estimate the parameters in  model (\ref{eqn1.1}),
\cite{carroll1997generalized} proposed a backfitting algorithm, which may lead to
unstable estimators. To deal with this issue, \cite{yu2002penalized}
introduced a penalized spline estimation procedure. \cite{xia2006semi}
developed an estimator based on the minimum average variance estimation.
\cite{liang2010estimation} proposed a profile least squares estimation
procedure. Under a mild assumption that the covariate $\bZ$ has a dimension
reduction structure on the covariate $\bX$, \cite{wang2010estimation} proposed
a two-stage procedure. \cite{zhu2006empirical} conducted confidence regions
for the parameters in  model (\ref{eqn1.1}) based on bias-corrected
empirical likelihood. However, these works only deal with the case when dimension of $\bX$ is small and fixed.



When the dimension of $\bX$ is large compared with the sample size
$n$, it is challenging  to estimate the parameters. To handle this
issue, it is natural to assume that the high dimensional nuisance
parameter $\bal$ is sparse. In other words, only a small number of
elements of $\bal$ are
nonzero. 
Under this sparsity assumption, variable selection procedures are
developed by many authors. See for instance,
\cite{liang2010estimation},  \cite{zhang2013partial},
\cite{lai2014estimation}, and \cite{zhang2017estimation}. However,
these studies all focus on the situation that $p$ and $q$ are both
fixed. As an exception, when the dimension $p+q=o(n^{1/3})$,
\cite{zhang2012dimension} considered the estimation and variable
selection problem for PLSIM. In this paper, we adopt penalty-based
variable selection methods such as \cite{tibshirani1996regression}
and \cite{fan2001variable} to ultrahigh dimensional PLSIM. A
notable feature of our approach is that we only penalize the
nuisance parameter $\bal$, and do not penalize the parameter of
interest $\bbeta$.

In addition to estimation and variable selection, we are also
interested in making statistical inference on the parametric
component $\bbeta$ and the nonparametric function $\eta(\cdot)$.
That is, to test whether $H_{01}: \bbeta=0$ and $H_{02}:
\eta(t)=g(t,\zeta)$ with $g(\cdot,\cdot)$ being known up to an
unknown parameter vector $\zeta\in\mathbb{R}^d$.
The hypothesis $H_{01}$ is usually called
significance testing problem, while the hypthesis $H_{02}$ is
referred to as model-specification testing problem. When the
dimension of $\bX$ is fixed, the specification testing
problem $H_{02}$ has been investigated by many authors. See for
instance, \cite{zheng1996consistent}, \cite{lavergne2012one},
\cite{guo2016model}, and \cite{li2016consistent}.
The tests for $H_{01}$ and $H_{02}$ may be
used to address the following interesting and important questions:
given $\bX$, does $\bZ$ carry additional information about the
response $Y$? Is the function form of $\eta(\cdot)$ linear or not?
 For these
questions, when the dimensions of $\bX$ and $\bZ$ are fixed,
\cite{liang2010estimation} constructed suitable test statistics
based on residual sums of squares under the null and alternative
hypotheses. However, the procedure developed in  \cite{liang2010estimation} are not
directly applicable for high dimensional $\bX$.
Moreover, due to the presence of nonparametric function
the setting studied in this paper is distinguished from the existing works on statistical inference
for the high dimensional parametric regression model.
To make inference for the
parameter $\bbeta$ in model (\ref{eqn1.1}), we have to estimate the nuisance
parameter $\bal$ and the nuisance function $\eta(\cdot)$. For high dimensional
$\bX$, this is very challenging. Without an appropriate estimation, high
dimensional nuisance parameter $\bal$ and the nuisance function $\eta(\cdot)$
may significantly deteriorate  the detection power of related testing methods.
We show that our proposed estimators are very helpful to make suitable inference
about the parametric component $\bbeta$ and the nonparametric function $\eta(\cdot)$.



This work makes several interesting contributions to the literature. Firstly, we establish
the asymptotic distributions of partial penalized least squares
estimator for high dimensional and even ultrahigh dimensional PLSIM.
Previous studies mainly focus on the finite and fixed dimension setting, while our procedure allows the dimensionality to be exponential order of the sample size and the sparsity level to be diverging. Secondly, we propose an F-type test for the
parametric component $\bbeta$ and show that the limiting null distribution is $\chi^2$ distribution.
Thirdly, we study the specification testing problem for the nonparametric
component $\eta(\cdot)$, propose a test statistic and show that it follows an asymptotic normal
distribution.


The paper is organized as follows. In section 2, we propose the partial
penalized least squares estimators and derive their asymptotic
distributions. In section 3, we propose tests for the parametric component
$\bbeta$ and the nonparametric part $\eta(\cdot)$, and derive their asymptotical
distributions. In section 4, numerical studies are conducted to illustrate the performances of our proposed
test statistics.  Conclusions and discussions are given in section 5. All proofs are given in the supplementary material of this paper.

\section{Profile partial penalized least squares estimators}


Suppose that $\{\bX_i, \bZ_i, Y_i\}, i= 1,\cdots, n$, is a sample from model
(\ref{eqn1.1}). To ensure model (\ref{eqn1.1}) identifiable, we assume
that $\|\bal\|_2=1$ and its first element is positive. This constraint reduces
dimension of $\bal$ from $p$ to $p-1$. As in \cite{yu2002penalized},
\cite{wang2010estimation} and \cite{cui2011efm}, we adopt the
`delete-one-component' method  and write
$\bal=((1-\|\bal^{(1)}\|^2_2)^{1/2},\bal^{(1)^\top})^\top $ with $\bal^{(1)}
=(\alpha_2,\cdots,\alpha_p)^\top $. Thus $\bal$ can be viewed as a function of
$\bal^{(1)}$, and the $p\times(p-1)$ Jacobian matrix is
\begin{eqnarray}\label{eqn2.1}
J(\bal^{(1)})=\frac{\partial\bal}{\partial \bal^{(1)}}=\left(
                                               \begin{array}{c}
                                                 -\dfrac{\bal^{(1)\top}}{(1-\|\bal^{(1)}\|^2_2)^{1/2}} \\
                                                 I_{(p-1)\times(p-1)} \\
                                               \end{array}
                                             \right).
\end{eqnarray}

Let $\btheta=(\bbeta^\top ,\bal^{(1)\top})^\top $, and denote
$\btheta_0=(\bbeta^\top _0,\bal^{(1)\top}_0)^\top $ to be the true value of $\btheta$.
Let $J_0=J(\bal^{(1)}_0)$. Further denote the estimator of $\btheta$ by
$\hat\btheta=(\hat\bbeta^\top ,\hat\bal^{(1)\top})^\top$, which will be specified later.

We next develop an estimation procedure for model (\ref{eqn1.1})  based on
profile least squares method. Specifically, for any given
$\btheta=(\bbeta^\top ,\bal^{(1)\top})^\top $, we use local linear regression to estimate
$\eta(\cdot)$ by minimizing
\begin{eqnarray}\label{eqn2.2}
\sum_{i=1}^n[Y_i-\bbeta^\top \bZ_i-d_0-d_1(\bal^\top\bX_i-t)]^2K_h(\bal^\top\bX_i-t),
\end{eqnarray}
with respect to $d_0$ and $d_1$, where $K_h(\cdot)=K(\cdot/h)/h$ is for a kernel
function $K(\cdot)$ with bandwidth $h$. Denote $(\hat d_0,\hat d_1)$ to be the
minimizer of (\ref{eqn2.2}). Then for given $\btheta$,
$\hat\eta(t;\btheta)=\hat d_0$ and $\hat\eta^{\prime}(t;\btheta)=\hat d_1$ for a
specific $t$. Specifically, define
\begin{eqnarray*}
S_{nl}(t;\bal)&=&\frac{1}{n}\sum_{i=1}^n(\bal^\top\bX_i-t)^lK_h(\bal^\top\bX_i-t);\\
U_{nk}(t;\bal)&=&K_h(\bal^\top \bX_k-t)\{S_{n2}(t;\bal)-(\bal^\top \bX_k-t)
S_{n1}(t;\bal)\};\\
W_{nj}(t;\bal)&=&U_{nj}(t;\bal)/\sum_{k=1}^n
U_{nk}(t;\bal).
\end{eqnarray*}
Then
\begin{equation}
\label{eqn2.3}
\hat\eta(\bal^\top\bX_i,\btheta)=\sum_{j=1}^n W_{nj}(\bal^\top\bX_i;\bal)
(Y_j-\bbeta^\top \bZ_j).
\end{equation}
Define partial penalized least squares function as
\begin{equation}\label{eqn2.4}
Q_n(\btheta,\lambda)=
\frac{1}{2n}\sum_{i=1}^n \{Y_i-\hat\eta(\bal^\top \bX_i,\btheta)-\bbeta^\top \bZ_i\}^2
+\sum_{j=1}^{p-1} p_{\lambda}(|\alpha^{(1)}_j|),
\end{equation}
where $p_{\lambda}(\cdot)$ is a penalty function with tuning parameter $\lambda$.
Minimizing (\ref{eqn2.4}) with respect to $\bbeta$ and $\bal^{(1)}$
leads to their estimates $\hat\bbeta$ and $\hat\bal^{(1)}$.
It is worth  noting that the nonparametric function $\eta(\cdot)$ is estimated locally,
while the parametric vectors $\hat\bbeta$ and $\hat\bal^{(1)}$ are obtained
globally by incorporating the penalty function. It is crucial that we penalize only the nuisance
parameter $\bal$. On one hand, it can significantly reduce the dimension of the nuisance parameter.
On the other hand, it would not shrink the small elements of
$\bbeta$ to be zero, and thus we can construct hypothesis testing on $\bbeta$ with local power.

\subsection{Theoretical results}
We next study the theoretical properties of the proposed estimation procedure.
Assume that the penalty function $p_{\lambda}(t_0)$ is increasing
and concave in $t_0\in [0,\infty)$, and has a continuous derivative $p'_{\lambda}(t_0)$ with
$p'_{\lambda}(0+)>0$. Let $\rho(t_0,\lambda)=p_{\lambda}(t_0)/\lambda$ for $\lambda>0$.
In addition, assume $\rho'(t_0,\lambda)$ is increasing in $\lambda\in(0,\infty)$ and
$\rho'(0+,\lambda)$ is independent of $\lambda$. For any vector $\mathbf{v}=(v_1,\cdots, v_r)^\top $, define
$$\bar{\rho}(\mathbf{v},\lambda)=\{\sgn(v_1)\rho'(|v_1|,\lambda),\cdots,\sgn(v_r)\rho'(|v_r|,\lambda)\}^\top,$$
where $\sgn(v_1)=I(v_1>0)-I(v_1<0)$. Following \cite{fan2011}, we further define the local concavity
of the penalty function $\rho$ at $\mathbf{v}$ as
$$\kappa(\rho,\mathbf{v},\lambda)=\lim_{\epsilon\rightarrow 0^+}\max_{1\leq j\leq r}
\sup_{t_1<t_2\in(|v_j|-\epsilon,|v_j|+\epsilon)}-\frac{\rho'(t_2,\lambda)-\rho'(t_1,\lambda)}{t_2-t_1}.$$

Before we proceed further, let us introduce some notations. Denote
$\mathcal{A}=\{j:\alpha^{(1)}_{0,j}\neq 0\}$ and $s=|\mathcal{A}|$ be the
number of elements in $\mathcal{A}$. Define
$\mathcal{A}^c=[1,\cdots,p-1]-\mathcal{A}$ be the complement set of
$\mathcal{A}$. Let $\bal^{(1)}_{\mathcal{A}}$ be the subvector of $\bal^{(1)}$
formed by elements in $\mathcal{A}$. Similarly let $B_{\A}$ be  the submatrix
of a matrix $B$ formed by columns in $\A$. Moreover $\bvarth=(\bbeta^\top
,\bal^{(1)\top}_{\mathcal{A}})^\top $, and $\bvarth_0, \hat\bvarth$ are
similarly defined. Further let $J_{\mathcal{A}}$ be the submatrix of $J$ formed
as follows
$$
J_{\mathcal{A}}(\bal^{(1)}_{\mathcal{A}})=\left(
\begin{array}{c}
                                                 -\dfrac{\bal^{(1)\top}_{\mathcal{A}}}{(1-\|\bal^{(1)}_{\mathcal{A}}\|^2_2)^{1/2}} \\
                                                 I_{s\times s} \\
                                               \end{array}
                                             \right).
$$
$J_{0,\mathcal{A}}=J_{\mathcal{A}}(\bal^{(1)}_{0,\mathcal{A}})$.
Let
$l_n=\min_{j\in\mathcal{A}}|\alpha^{(1)}_{0,j}|/2$, the half minimum
signal of $\bal^{(1)}_{0,\mathcal{A}}$. Define
$\mathcal{N}_0=\{\btheta\in\mathbb{R}^{p+q-1}:\|\bvarth-\bvarth_0\|_2\leq l_n\}$.
Let $\Gamma_{i0}=\bal_0^\top\bX_i, \Gamma_{i1}=\bal^\top\bX_i, \eta_{i0}=\eta(\Gamma_{i0})$, and $\eta_{i1}=\eta(\Gamma_{i1})$. Further
let  $\A^*=\A\cup\{1\}, \mu_{1}(\Gamma_{i1})=E[\bX_{i,\A^*}|\Gamma_{i1}], \widetilde \bX_i(\bal)=
\bX_{i,\A^*}-\mu_{1}(\Gamma_{i1}), \mu_{2}(\Gamma_{i1})=E[\bZ_i|\Gamma_{i1}]$, $\widetilde \bZ_i(\bal)=\bZ_i-\mu_{2}(\Gamma_{i1})$, $\mu^{\star}_{1i}=\mu_1(\Gamma_{i0}), \mu^{\star}_{2i}=\mu_2(\Gamma_{i0}), \widetilde \bX^{\star}_i=\widetilde \bX_i(\bal_0),$ $\widetilde \bZ^{\star}_i=\widetilde \bZ_i(\bal_0)$, and $c_n=O(h^2+\sqrt{\log n/nh^3})$. Define $L_{i}=\left(\begin{array}{c}
\widetilde{\mathbf{Z}}_{i}^{\star} \\
\eta_{i 0}^{\prime} J_{0, \mathcal{A}}^{\top}{\mathbf{X}}_{i}^{\star}
\end{array}\right)$, $\Sigma^{\star}=E[L_iL_i^T].$
For a vector $\mathbf v, \|\mathbf v\|_{\infty}=\max|v_i|$. For matrix $B$, denote $\lambda_{\min}(B)$ and $\lambda_{\max}(B)$ to be the minimum and maximum eigenvalues of the matrix $B$. $\|B\|_{2,\infty}=\sup_{\bm{v}:\|\bm{v}\|_2=1}\|B\bm{v}\|_{\infty}$.
Throughout the paper, $c$ and $C$ are two generic positive constants.

We impose the following conditions:

\begin{itemize}
\item [(A1)] $\lambda_{\min}(\Sigma^{\star})\geq c>0$, $\lambda_{\max}(\Sigma^{\star})=O(1)$. For any $1\leq i\leq n, 1\leq j\leq p$, $X_{ij}$ are uniformly bounded.
\item [(A2)]$l_n\gg\lambda_{n}\gg\max\{n^{1/\varpi+\varsigma}c_n, n^{1/\varpi+\varsigma}\sqrt{s/n}, \sqrt{\log p/n}\}$ with some $\varpi\geq 8$ and some arbitrary small $\varsigma>0$, $p'_{\lambda_n}(l_n)=o((ns)^{-1/2})$, $\lambda_n\kappa_0=o(1)$ where
$\kappa_0=\max_{\bdelta\in\mathcal{N}_0}\kappa(\rho,\bdelta,\lambda_{n})$. Here $a\gg b$ means $\lim_{n\rightarrow \infty} a/b=\infty$.

\item [(A3)] Assume that $E(|\epsilon_i|^{\varpi})\leq C$ and ${\log p}/{n^{1-2/\varpi-\varsigma}}\rightarrow 0$. For any $\bal$ with $\|\bal-\bal_0\|_2=O(\sqrt{s/n})$, let $R_{ki}=(J^\top(\bX_i-E[\bX_i|\bX^\top_i\bal]))_k\eta'_{i1}$. Let $\bR=(R_{ki})_{i=1,\cdots,n}^{k=1,\cdots,p}$ and $\tilde{\bZ}(\bal)=(\tilde \bZ_1(\bal),\cdots,\tilde \bZ_n(\bal))$. Assume that $$\|\bR^{T}_{\mathcal{A}^c}(\tilde{\bZ}(\bal), \bR_{\mathcal{A}})\|_{2,\infty}=O_p(n^{1+1/\varpi+\varsigma}),\,\,\,\max_{1\leq l\leq q}\sup_{\btheta\in\mathcal{N}_0}E[\tilde Z_l^4(\bal)|\Gamma_{i1}]<C<\infty.$$
$E(L_1^TL_1|\Gamma_{10}=\gamma)/s$ is bounded uniformly for $\gamma$.

\item [(A4)]$\eta(\cdot)$ is twice order continuously differentiable on its support $\mathcal{D}$. $\eta'(\cdot)$ and $\eta''(\cdot)$ are bounded on $\mathcal{D}$.  For any $\bal$ with $\|\bal-\bal_0\|_2=O(\sqrt{s/n})$, the density function of $\bal^\top \bX$, $f(x)$ is bounded away from 0 on $\mathcal{D}$. $\sup_x f(x)\leq B_0<\infty$. For some $q>0 ,\delta>0$ and $M>2$, $E|\Gamma_{i1}|^{2q+\delta}<\infty$.
If $T_{ik}$ is one of the following: $X_{ik}, X_{ik}\eta_{i1},X_{ik}\eta'_{i1},$
\[\sup_{k\in \mathcal{A}}\sup_xE(|T_{1k}|^M|\Gamma_{11}=x)f(x)\leq B_1<\infty,\]
\[\sup_{k\in \mathcal{A}}\sup_x|x|^qE(T_{1k}|\Gamma_{11}=x)f(x)\leq B_2<\infty.\]
 First order derivatives of
 $f(x)\mu_{1i}(x),f(x)\mu_{2j}, f(x)\mu_{1i}(x)\eta(x), f(x)\eta(x)$ exist and are L-Lipschitz, and  $\eta^\prime(x)\mu_{1i}(x),\mu_{2j}(x)$ are L-Lipschitz, where $\mu_{1i}(x)$, $\mu_{2j}(x)$ are the $i$th and $j$th components of $\mu_{1}(x)$, $\mu_{2}(x)$ respectively, $i\in\mathcal{A}^\star$, $1\leq j\leq q$, $0<L<\infty$.

\item [(A5)] Suppose $K(u)$ is a differentiable and symmetric kernel function and $G(u)$ is one of the following:  $K(u), tK(u), t^2K(u), tK'(u), t^2K'(u)$. $G(u)$ is  satisfying that
$|G^\prime(u)|\leq \lambda_1<\infty$. When $|u|>L$,  $|G^\prime(u)|\leq \lambda_1|x|^{-\eta}$, $|G(u)|\leq \lambda_2|u|^{-q}\leq\bar{G}<\infty$ for some $\eta>1$, $L>0$ and $\lambda_2<\infty$. $\int|G^2(u)|du<\infty.$ $\int u^4|G^2(u)|du<\infty.$

\item [(A6a)] Assume that $s=o(n^{1/2})$, $h\log n \rightarrow 0$, $nh^8\rightarrow 0$, $nh^3/(s\log n)\rightarrow \infty,$ and \\$nh^4/(\log n)^2\rightarrow \infty$.
\item [(A6b)] Assume that $s=o(n^{1/3}), s^{3}h^4\rightarrow 0, sh\log n\rightarrow 0, nsh^8\rightarrow 0,$ $nh^3/(s^3\log n)\rightarrow \infty,$  and $nh^4/(s(\log n)^2)\rightarrow \infty$.
\end{itemize}


These conditions are mild and commonly assumed. The uniform
boundedness of elements of covariate $\bX$ in condition (A1) is
usually assumed to facilitate the technical arguments, see for
instance \cite{wang2010estimation} and
\cite{sherwood2016partially}. But for $\bZ$, uniform boundedness
is not required. The effect of nonparametric estimation is taken
into account in Condition (A2). The rates of $l_n$ and $\lambda_n$
are required for sparsity. In condition (A2), a minimum signal
assumption is imposed on the nuisance parameter $\bal$. This is
required for variable selection consistency and evaluation of the
uncertainty of the estimation for small signals. However, we
should emphasize that no minimum signal condition is imposed on
$\bbeta$, the parameter of primary interest. Thus, we can still
detect local alternative hypotheses effectively. This condition is
reasonable in many practical applications.
\cite{van2014asymptotically} and \cite{ning2017general} do not
impose such conditions. However, some additional assumptions are
imposed. In fact, the validity of the decorrelated score statistic
depends on the sparsity of additional parameter $\mathbf{w}^*$.
From Remark 6 in \cite{ning2017general}, we know that for testing
univariate parameters of high dimensional linear regression model,
this requires the degree of a particular node in the graph to be
relatively small when the covariate follows a Gaussian graphical
model. A further discussion about the decorrelated score statistic
is given in Remark 1. From Conditions (A2) and (A3), clearly the
dimensionality $p$ is allowed to be exponential order of the
sample size $n$. In the theory of high dimensional regression,
Gaussian or sub-Gaussian tail condition for the random error
$\epsilon$ is usually assumed. However, in this paper, we only
require a very mild moment condition. Condition (A4) is some
regularity conditions for uniform convergence for kernel
estimation and for the smoothness of related functions. Condition
(A5) is the usual condition for the kernel function $K(\cdot)$ and
is satisfied when $K(\cdot)$ is the density of normal distribution
or a smooth density function with compact support.

Condition (A6a) is required to control the estimation
error due to nonparametric estimators, while (A6b) is stronger and
needed for asymptotic representation of $\hat\btheta$.  If we set $s$
to be fixed, then conditions (A6a) and (A6b) reduce to be
$nh^3/\log n\rightarrow \infty, nh^4/(\log n)^2\rightarrow \infty,nh^8\rightarrow 0$ and $h\log n \rightarrow 0$, which are assumed by
many authors for partially linear single-index model with fixed
dimension. Thus conditions (A6a) and (A6b) are
modifications of classical conditions to accounting for the effect of $s$. 
Take $h=O(n^{-c_1}), s=O(n^{c_2}), 1/8<c_1<1/4, 0\leq c_2<1/2$, then
conditions in (A6a) are satisfied when
$1-3c_1-c_2>0$. This leads to $c_2<1-3c_1.$ If we set $c_1=1/5$,
we can get $0\leq c_2<2/5$. For condition (A6b), it requires that
$$
c_2<1/3,\,\,c_2<c_1,\,\,1+c_2-8c_1<0, \,\,1-3c_1-3c_2>0,1-4c_1-c_2>0.
$$
These lead to
$$
c_2<\min\{\frac{1}{3},c_1,8c_1-1,\frac{1}{3}-c_1, 1-4c_1 \}.
$$
If we set $c_1=1/5$, the condition is satisfied provided that $0\leq c_2<2/15$.

We first establish the rate of convergence of $\hat{\bal}$  and its sparsity, and then
derive an asymptotic representation of $\hat\bvarth$ in the following theorem.
\begin{theorem}
\label{thm1} Under Conditions (A1)-(A6a), the following holds. With probability
tending to 1, $\hat\bal^{(1)}$ must satisfy (i)
$\hat\bal^{(1)}_{\mathcal{A}^c}=0$. (ii)
$\|\hat\bal^{(1)}_{\mathcal{A}}-\bal^{(1)}_{0,\mathcal{A}}\|_2=O_p(\sqrt{s/n})$.
If further condition (A6b) holds, we obtain that
\begin{eqnarray}\label{eqn2.5}
\sqrt n(\hat\bvarth-\bvarth_0)&=&\Sigma^{\star -1}\frac{1}{\sqrt n}\sum_{i=1}^n \epsilon_i\left(
  \begin{array}{ccc}
    \widetilde \bZ^{\star}_i\\
   \eta'_{i0}J^\top _{0,\mathcal{A}}\widetilde \bX_i^{\star}\\
  \end{array}
\right)+o_p(1).
\end{eqnarray}
\end{theorem}
The proof of this theorem is given in the supplementary material of this paper.
Theorem 1 establishes the sparsity, consistency and asymptotic representation of the proposed profile partial penalized least squares estimators. Our results allow the dimensionality $p$ to be exponential order of the sample size $n$. The sparsity level $s$ is also allowed to be diverging. Theorem~\ref{thm1} implies that the optimal bandwidth $h=O(n^{-1/5})$
may be used for not only the sparsity and consistency,
but also for the asymptotic representation, and $s$ may be of order $o(n^{2/5})$
for the sparsity and consistency. However, for the asymptotic representation,
more restrictive conditions on $h$ and $s$ as in condition (A6b) are required.

\section{Hypothesis testing}

This section aims for developing hypothesis testing procedures for both $\bbeta$
and $\eta(\cdot)$.

\subsection{Testing the parametric component}
Of interest is to test hypothesis
\begin{equation}
H_{01}: \bbeta=0 \quad \mbox{versus}\quad H_{11}: \bbeta\ne 0.
\label{eqn3.1a}
\end{equation}
Under $H_{11}$, the residual sum of squares is
\begin{equation}\label{eqn3.1}
    \mbox{RSS}_1=\sum_{i=1}^n [Y_i-\hat\eta(\hat{\bal}^\top \bX_i,\hat\btheta)-\hat\bbeta^\top  \bZ_i]^2.
\end{equation}
Under $H_{01}$, we need to consider the constrained penalized least squares estimators. Specially, for any given
$\btheta=(\widetilde\bbeta^\top , \bal^{(1)\top})^\top $ with $\widetilde\bbeta=0$,
we first obtain the constrained profile estimator for the nonparametric function:
\begin{equation}
\label{eqn3.2}
\widetilde\eta(\bal^\top \bX_i,\btheta)=\sum_{j=1}^n W_{nj}(\bal^\top\bX_i;\bal)
Y_j.
\end{equation}
Denote the constrained penalized least squares function as
\begin{equation}\label{eqn3.3}\widetilde Q_n(\btheta,\lambda)=\frac{1}{2n}\sum_{i=1}^n [Y_i-\widetilde\eta(\bal^\top \bX_i,\btheta)]^2+\sum_{j=1}^{p-1} p_{\lambda}(|\alpha^{(1)}_j|),\end{equation}
for some penalty function $p_{\lambda}(\cdot)$ with a tuning parameter $\lambda$. Minimizing the above objective function with respect to $\bal^{(1)}$ leads to the estimator $\widetilde\bal^{(1)}$. Afterwards, we define the residual sum of squares $RSS_0$
under the null hypothesis:
\begin{equation}\label{eqn3.4}
    \mbox{RSS}_0=\sum_{i=1}^n [Y_i-\widetilde\eta(\widetilde{\bal}^\top\bX_i,\widetilde\btheta)]^2.
\end{equation}
Here $\widetilde\btheta=(0^\top ,\widetilde\bal^{(1)\top})^\top $. Under the null hypothesis, $RSS_0$ and $RSS_1$ should be close, while under the alternative hypothesis, $RSS_0$ should be larger than $RSS_1$. This motivates us to
consider the following test statistic:
\begin{equation}\label{eqn3.5}
T_n=\frac{RSS_0-RSS_1}{RSS_1/(n-q)}.
\end{equation}

We impose the following condition.
\begin{itemize}
\item[(A7)] Assume that $\|\bm{\delta}_n\|_2=O(\sqrt{1/n})$. Here $\bm{\delta}_n$ corresponds to local alternative
hypotheses $H_{11}^{(n)}: \bbeta=\bm{\delta}_n$.
\end{itemize}

\begin{theorem}\label{thm2}
Suppose that Conditions (A1)-(A7) hold, then we have
\begin{eqnarray}\label{eqn3.6}
\sup_x|P(T_n\leq x)-P(\chi^2_{q}(n\bm{\delta}_n^\top \Phi^{-1}\bm{\delta}_n/\sigma^2)\leq x)|\rightarrow 0.
\end{eqnarray}
where $\Phi=(I_q, 0_{q\times s})\Sigma^{\star -1}(I_q, 0_{q\times s})^\top $
and $\chi^2_{q}(n\bm{\delta}_n^\top \Phi^{-1}\bm{\delta}_n/\sigma^2)$ is
a noncentral chi-square random variable with $q$ degrees of freedom and noncentrality parameter $n\bm{\delta}_n^\top \Phi^{-1}\bm{\delta}_n/\sigma^2$ which is allowed to vary with $n$.
\end{theorem}

The proof of Theorem~\ref{thm2} is given in the supplementary material of this paper.
From Theorem \ref{thm2}, it is clear that the null distribution of $T_n$ is a chi-square random variable with $q$ degrees of freedom, which does not depend on nuisance parameter $\bal$ nor the nuisance function $\eta(\cdot)$. This implies that the so-called Wilks phenomenon still holds even in the (ultra)high dimensional partially linear single-index model. Further, the test statistic $T_n$ can detect local alternatives, which converge to the null hypothesis at the root-$n$ rate.

\begin{remark}\em
In a case that the minimum signal assumption in (A2)
is believed to be invalid, we may want to extend the decorrelated
score statistic to make inference about $\bbeta$. However, this is
not straightforward. For high dimensional partially linear single
index models, the negative Gaussian quasi-loglikelihood (i.e., the
least squares loss) is
$$l(\bbeta,\bal)=\frac{1}{n}\sum_{i=1}^n [Y_i-\bbeta^\top
\bZ_i-\eta(\bal^\top \bX_i)]^2.$$ By adopting the notations in
\cite{ning2017general}, we may consider the score functions
$\nabla_{\bbeta}l(\bbeta,\bal)=n^{-1}\sum_{i=1}^n [Y_i-\bbeta^\top
\bZ_i-\eta(\bal^\top \bX_i)]\bZ_i$ and
$\nabla_{\bal}l(\bbeta,\bal)=n^{-1}\sum_{i=1}^n [Y_i-\bbeta^\top
\bZ_i-\eta(\bal^\top \bX_i)] \eta'(\bal^\top
\bX_i)J(\bal^{(1)})^\top \bX_i$.  When the dimension of $\bX$ is
high, we cannot directly use $\nabla_{\bbeta}l(\bbeta,\bal)$ to
make inference for the  parameter of interest $\bbeta$. Instead we
let
$$S(\bbeta,\bal)=\nabla_{\bbeta}l(\bbeta,\bal)-\mathbf{W}^\top \nabla_{\bal}l(\bbeta,\bal).$$ Here $\mathbf{W}^\top=E[\nabla_{\bbeta}l(\bbeta,\bal)\nabla^\top_{\bal}l(\bbeta,\bal)]E[\nabla_{\bal}l(\bbeta,\bal)\nabla^\top_{\bal}l(\bbeta,\bal)]^{-1}
\in\mathbb{R}^{q\times (p-1)}$. Immediately, it follows that
$E[S(\bbeta,\bal)\nabla^\top_{\bal}l(\bbeta,\bal)]=0$. That is,
$S(\bbeta,\bal)$ is uncorrelated with
$\nabla_{\bal}l(\bbeta,\bal)$. We then regard $S(\bbeta,\bal)$ as
the decorrelated score function for $\bbeta$. The key idea is to
apply a high dimensional projection of the score function of the
interested parameter to the nuisance parameter space.

To make inference about $\bbeta$ based on $S(\bbeta,\bal)$, we need to estimate the nuisance parameter $\bal$, the unknown matrix $\mathbf{W}$ and the unknown functions $\eta(\cdot)$ and $\eta'(\cdot)$. For $\bal$, $\eta(\cdot), \eta'(\cdot)$ and $J(\bal^{(1)})$, we may adopt the partial penalized least squares estimators introduced in this paper. While for the estimator of $\mathbf{W}$, we can consider the following formula:
$$\hat{\mathbf{W}}=\arg\min_{\mathbf{W}}\frac{1}{2n}\sum_{i=1}^n \|\nabla_{\bbeta}l(\hat\btheta)-\mathbf{W}^\top\nabla_{\bal}l(\hat\btheta)\|^2_2+p_{\lambda'}(\mathbf{W}),$$
for some penalty function $p_{\lambda'}(\cdot)$ with tuning parameter $\lambda'$. Besides the sparsity of $\bal$, we also require $\mathbf{W}$ to be sparse. The estimated de-correlated score function is:
$$\hat S(0, \hat\bal)=\nabla_{\bbeta}l(0,\hat\bal)-\hat{\mathbf{W}}^\top\nabla_{\bal}l(0,\hat\bal).$$

To prove the asymptotic normality of $\hat S(0, \hat\bal)$, it is
required to check the assumptions 3.1-3.4 in Ning and Liu (2017).
Due to the complicated formula of $\hat S(0, \hat\bal)$, these
four assumptions are not easy to verify. The investigation of the
asymptotic behavior of the decorrelated score statistic is beyond
the scope of this work.
\end{remark}
\subsection{Testing the nonparametric component}

In practice, it is of interest in testing whether the nonparametric
function $\eta(\cdot)$ is in a specific form, for example linear, or not, or even whether it is  constant.
This motivates us to consider the following null hypothesis
$$
H_{02}: \eta(t)\equiv g(t, \zeta) \quad\mbox{versus}\quad
H_{12}: \eta(t)\ne g(t, \zeta) \ \mbox{for\ some} \ t,
$$
where the form $g(\cdot,\cdot)$ is known, and $\zeta\in \mathbb{R}^d$
with $d$ being fixed is an unknown parametric vector. For example, if
we set $g(t, \zeta)=c$, a constant, then  $H_{02}$ corresponds to
testing whether the predictor $\bX$ contributes to the response $Y$ or not. When the dimension of $\bX$ is fixed, this kind of specification testing
problem has been investigated by many authors. See for instance, \cite{zheng1996consistent}, \cite{guo2016model}, and \cite{li2016consistent}.

Let $\epsilon_{0i}=Y_i-\bbeta^\top _0\bZ_i-g(\bal^\top _0 \bX_i,\zeta_0)
=\eta(\bal^\top _0 \bX_i)-g(\bal^\top _0 \bX_i,\zeta_0)+\epsilon_i$.
Then it follows that $E(\epsilon_{0i}|\bal^\top _0 \bX_i)=\eta(\bal^\top _0 \bX_i)
-g(\bal^\top _0 \bX_i,\zeta_0)$. Clearly, under $H_{02}$,
$E(\epsilon_{0i}|\bal^\top _0 \bX_i)=0$, while under $H_{12}$, it is not equal to zero.
Further we have under $H_{02}$, $E[\epsilon_{0i}E(\epsilon_{0i}|\bal^\top _0\bX_i)f(\bal^\top _0\bX_i)]
=0$, while under $H_{12}$, we have
$$E
[\epsilon_{0i}E(\epsilon_{0i}|\bal^\top _0\bX_i)f(\bal^\top _0 \bX_i)]\neq 0.
$$
This motivates us to propose the following test statistic
\begin{eqnarray}\label{eqn3.7}
S_n=\frac{1}{n(n-1)}\sum_{i=1}^n\sum_{j\neq i}^n\hat\epsilon_{0i}\hat\epsilon_{0j}
\frac{1}{b}G(\frac{\hat\bal^\top (\bX_i-\bX_j)}{b}),
\end{eqnarray}
where $G(\cdot)$ is a kernel function, $b$ is the bandwidth,
$\hat\epsilon_{0i}=Y_i-\hat\bbeta^\top \bZ_i-g(\hat\bal^\top  \bX_i,\hat\zeta)$,
$\hat\bbeta,\hat\bal$ are the partial penalized least
squares estimators defined in section 2, and $\hat\zeta$ is the nonlinear
least squares estimator obtained by minimizing
$$
\min_{\zeta}\frac{1}{n}\sum_{i=1}^n[Y_i-\hat\bbeta^\top \bZ_i-g(\hat\bal^\top  \bX_i,\zeta)]^2.
$$

Denote
\begin{eqnarray*}
&&\sigma^2_S=2\int G^2(t)dt\cdot\int \sigma^4 f^2(\bal^\top _0 \bX)d(\bal^\top _0 \bX);\\
&&\hat\sigma^2_S=\frac{2}{n(n-1)}\sum_{i=1}^n\sum_{j\neq i}^n
\frac{1}{b}G^2(\frac{\hat\bal^\top (\bX_i-\bX_j)}{b})\hat\epsilon^2_{0i}\hat\epsilon^2_{0j};\\
&&g_{01}(\cdot,\zeta_0)=\partial g(\cdot,\zeta)/\partial \zeta|_{\zeta=\zeta_0};\\ &&g_{k0}(\bal^\top _0\bX_i,\cdot) =\partial^k g(\bal^\top \bX_i,\cdot)/\partial^k (\bal^\top \bX_i)|_{\bal=\bal_0},k=1,2;\\
&&\bN_i=(\bZ_i^\top , g_{10}(\bal^\top _0\bX_i,\zeta_0)\bX^\top _{i,\mathcal{A}})^\top ;\,\,\, \bM_i=(\bN^\top _i, g^\top _{01}(\bal^\top _0\bX_i,\zeta_0))^\top .
\end{eqnarray*}

The following conditions are needed to facilitate the technical proofs.
\begin{itemize}
\item[(B1)] The kernel function $G(\cdot)$ is univariate bounded, continuous and symmetric density function satisfying
$\int t^2G(t)dt<\infty$, and $\int |t|^j G(t)dt<\infty$ for $j=1, 2, 3$. The second order derivative of $G(\cdot)$ is bounded.
\item[(B2)] $\lambda_{\max}(E[\bM_i\bM^\top _i])<\infty$.
\item[(B3)] $g_{20}(\bal^\top \bX_i,\cdot)$ is bounded and $g_{01}(\bal^\top \bX_i,\zeta_0)$ satisfies the first order Lipschitz condition  for $\bal^\top \bX_i$ in a neighborhood of $\bal^\top _0 \bX_i$. Further assume that $E[g^2_{01}(\bal^\top_0 \bX_i,\zeta_0)]<\infty$.
\end{itemize}

We then have the asymptotic normality of the proposed test statistic.
\begin{theorem}\label{thm3}
Suppose that Conditions (A1)-(A6a) and (B1)-(B3) hold, under $H_{02}$
with conditions  $nb^2/s^2\rightarrow \infty$
and $sb^{1/2}\rightarrow 0$, we obtain that
\begin{eqnarray}
\label{eqn3.8}
nb^{1/2}S_n\rightarrow N(0,\sigma^2_S).
\end{eqnarray}
\end{theorem}

From the proof of Theorem~\ref{thm3} given in the supplementary material, we can find that it is not necessary to assume
homoscedasticity in order
to derive the asymptotic distribution of $S_n$.
In fact, even under heteroscedasticity, it can be obtained that
$\|\hat\bal-\bal\|_2=O_p(\sqrt{s/n})$. This may further induce the
results in Theorem~\ref{thm3}. If $s$ is set to be fixed,
the conditions about the bandwidth $b$ boil down to $nb^2\rightarrow\infty$
and $b\rightarrow 0$, which are very mild. On the other hand,
if $s$ also diverges, restrictions are necessary.
That is, we need $s=o(\min\{b^{-1/2}, (nb^2)^{1/2}\})$.

We now standardize $S_n$ to obtain a scale-invariant statistic:
$$V_n=\sqrt{\dfrac{n-1}{n}}\dfrac{nb^{1/2}S_n}{\sqrt{\hat\sigma^2_S}}.$$

\begin{corollary}
Suppose that Conditions (A1)-(A6a) and (B1-B3) hold, under the null hypothesis with conditions  $nb^2/s^2\rightarrow \infty$ and $sb^{1/2}\rightarrow 0$, we obtain that
\begin{eqnarray}\label{eqn3.9}V^2_n\rightarrow \chi^2_1.\end{eqnarray}
Here $\chi^2_1$ is the chi-square distribution with one degree of freedom.
\end{corollary}

\subsection{Practical implementation issues}

In practice, it is desirable to have a data-driven method to choose
the tuning parameter $\lambda$.
%
We modify the high-dimensional BIC-type (HBIC) criterion proposed
by \cite{wang2013calibrating} to select  $\lambda$ by minimizing
$$
\text{HBIC}(\lambda)=\log(\hat{\sigma}^2_\lambda)+|\mathcal{A}_\lambda|\dfrac{C_n\log p}
{ n},
$$
where $\mathcal{A}_\lambda = \{j :\hat{\alpha}_j^{(1)}\neq 0\}$ is
the model identified by minimizing (\ref{eqn2.4}),
\[
\widehat{\sigma}_{\lambda}^{2}=\dfrac{1}{n}\sum_{i=1}^{n}\left[Y_{i}
-\widehat{\eta}\left(\hat{\bal}^\top\bX_{i},
\widehat{\boldsymbol{\theta}}\right)-\widehat{\boldsymbol{\beta}}^{\top} \bZ_{i}\right]^{2},
\]
and $C_n$ is a sequence of numbers that should diverge to $\infty$ slowly.
\cite{wang2013calibrating} suggested setting $C_n=\log(\log n)$.
This works well in a variety of settings in this paper.

We next discuss how to make minimization of (\ref{eqn2.4}) and (\ref{eqn3.3})
faster by using local linear approximation. Minimization problems of (\ref{eqn2.4})
and (\ref{eqn3.3}) are similar, so we only work on (\ref{eqn2.4}) as an example.
To minimize (\ref{eqn2.4}), noticing that a local linear approximation of
$\widehat{\eta}\left(\breve{\bal}^\top\bX_{i}, \breve{\boldsymbol{\theta}}\right)$ is
$$
\hat{\eta}(\breve{\bal}^\top\bX_i,\boldsymbol{\breve\theta})
\approx \hat{\eta}(\bal^\top\bX_i,\boldsymbol{\theta})+\left.
\frac{\partial \hat{\eta}(\bal^\top\bX_i,\boldsymbol{\theta})}
{\partial\left(\boldsymbol{\alpha}^{(1)\top},\boldsymbol{\beta}^\top\right)}
\right|_{(\boldsymbol{\alpha}^{(1)},\boldsymbol{\beta})}
\left(\begin{array}{c}
{\breve{\boldsymbol{\alpha}}^{(1)}-\boldsymbol{\alpha}^{(1)}}\\
{\breve{\boldsymbol{\beta}}-\boldsymbol{\beta}}\end{array}\right).
$$
$Q_{n}(\breve{\boldsymbol{\theta}}, \lambda)$ can be approximated by
\begin{equation}\label{eqn3.10}
Q_{n}(\breve{\boldsymbol{\theta}}, \lambda)\approx Q^L_{n}(\breve{\boldsymbol{\theta}},\boldsymbol{\theta}, \lambda):=\frac{1}{2n}\sum_{i=1}^{n}\left(Y_{i}^{*}-\bZ_{i}^{*\top}\breve{\boldsymbol{\theta}}\right)^{2}+\sum_{j=1}^{p-1} p_{\lambda}\left(|\breve{\alpha}_{j}^{(1)}|\right),
\end{equation}
where
$$Y_{i}^{*}=Y_i-\widehat{\eta}\left(\bal^\top\bX_{i},\boldsymbol{\theta}\right)+
\left.\frac{\partial\hat{\eta}(\bal^\top\bX_i,\boldsymbol{\theta})}{\partial\left( \boldsymbol{\alpha}^{(1)\top },\boldsymbol{\beta}^\top \right)}\right|_{\left( \boldsymbol{\alpha}^{(1)},\boldsymbol{\beta}\right)}
\left(\begin{array}{c}  {\boldsymbol{\alpha}^{(1)}}
\\
{\boldsymbol{\beta}}\end{array}\right),
$$
and
$$
\bZ_{i}^{*}=\left\{\left.\frac{\partial\hat{\eta}(\bal^\top\bX_i,\boldsymbol{\theta})}{\partial\left( \boldsymbol{\alpha}^{(1)\top },\boldsymbol{\beta}^\top \right)}\right|_{\left( \boldsymbol{\alpha}^{(1)},\boldsymbol{\beta}\right)}\right\}^\top+\left(\begin{array}{c} \boldsymbol{0}_{(p-1)\times1}\\{\bZ_i}\end{array}\right).
$$

We can solve (\ref{eqn2.4}) by iteratively minimizing penalized least squares functions.
Specifically, we could take the LASSO estimate for the whole model as the initial
value $\boldsymbol{\theta}^0$: For $k=1,2,...,$ we iteratively solve
(\ref{eqn3.11}) until the sequence of $\{\boldsymbol{\theta}^k\}$ converges. Our numerical study shows that the algorithm can converge very fast even if the initial value is not taken close to the true value.
\begin{equation}\label{eqn3.11}
\boldsymbol{\theta}^{k+1}=\arg\min_{\boldsymbol{\theta}} Q^L_{n}(\boldsymbol{\theta},\boldsymbol{\theta}^{k}, \lambda).
\end{equation}

\cite{zou2008one} proposed an algorithm for maximizing the penalized
likelihood for concave penalty functions based on local linear
approximation (LLA). Here we may apply local linear approximation to
the penalty term in (\ref{eqn3.10}),
\begin{equation}\label{eqn3.12}
p_{\lambda}\left(|\breve{\alpha}^{(1)}_{j}|\right) \approx p_{\lambda}\left(|\alpha^{(1)}_{j}|\right)+p_{\lambda}^{\prime}\left(|\alpha_{j}^{(1)}|\right)\left(|\breve{\alpha}_{j}^{(1)}|-|\alpha_{j}^{(1)}|\right),  \text { for } \breve{\alpha}^{(1)}_{j} \approx \alpha^{(1)}_{j}.
\end{equation}

Plugging (\ref{eqn3.12}) into (\ref{eqn3.10}) with constant terms omitted, we have
\begin{equation}\label{eqn3.13}
 Q^P_{n}(\breve{\boldsymbol{\theta}},\boldsymbol{\theta}, \lambda)=\frac{1}{2n}\sum_{i=1}^{n}\left(Y_{i}^{*}-\bZ_{i}^{*\top}\breve{\boldsymbol{\theta}}\right)^{2}+\sum_{j=1}^{p-1} p^{\prime}_{\lambda}(|\alpha_{j}^{(1)}|)(|\breve{\alpha}^{(1)}_j|).
\end{equation}

The minimization problem (\ref{eqn3.11}) becomes
\begin{equation}\label{eqn3.14}
\boldsymbol{\theta}^{k+1}=\arg\min_{\boldsymbol{\theta}} Q^P_{n}(\boldsymbol{\theta},\boldsymbol{\theta}^{k}, \lambda).
\end{equation}

In that way, we transform the original problem into iteratively solving penalized least
squares  with $L_1$ penalty. There are effective algorithms for
solving (\ref{eqn3.14}) because  dealing with $L_1$ penalty can take advantage
of kinds of computationally efficient  algorithms for the LASSO,
such as the least angle regression (LARS) algorithm proposed by \cite{efron2004least}.

\section{Numerical studies}
\subsection{Simulation studies}
In this section, we conduct simulation studies  to assess the finite-sample
performance of the proposed estimation and testing methods. The SCAD penalty function (Fan and Li, 2001)
is adopted in our simulation study. We also compare the proposed estimator and tests with the
oracle estimator and oracle tests, where the true signal set $\mathcal{A} =\{j: \alpha^{(1)}_{0,j}\neq 0\}$ is
assumed to be known and models are fitted based on that. Denote $T^O_n$
and $V^O_n$ to be the oracle test statistics. We report the performances of the profile partial penalized least squares estimator and different test statistics based on
500 replications. The sample size is taken to be 200. All simulations were conducted using MATLAB code.

\textbf{Example 1}. To evaluate the performance of the proposed profile partial penalized least squares estimator and  F-type test statistic
$T_n$, we generate simulation data from the following models.

\textit{Model 1a.}
$Y=\exp(\boldsymbol{\alpha}^\top\bX)+\boldsymbol{\beta}^\top\bZ+\epsilon$,

\textit{Model 1b.}
$Y=\sin(\dfrac{\pi}{2}\boldsymbol{\alpha}^\top\bX)+\boldsymbol{\beta}^\top\bZ+0.5\epsilon$.


For both models, we take the dimension of $\boldsymbol{\beta}$ to be  $2$ and
generate $\epsilon$ from  $N(0,1)$. In model 1a, the coefficient $\boldsymbol{\alpha}$ for
the mean function is $(\alpha_1,\alpha_2,\alpha_{10})^\top
=(2,1,1)/\sqrt6$ and $\alpha_j=0$ when $j\neq 1, 2, 10$. We generate $\bX= (X_1, X_2, \dots, X_p)^\top $
and $\bZ=(Z_1,Z_2)^\top $ from multivariate normal distributions with zero mean
and covariance matrix $\boldsymbol{\Sigma}_\bX=\left(\sigma_{i j}\right)_{p
\times p}$ and $\boldsymbol{\Sigma}_\bZ=\left(\sigma_{i j}\right)_{2 \times 2}$
respectively, with $\sigma_{i j}=0.25^{|i-j|}$. We consider  two scenarios for
the dimension of $\bX$, $p$: (i) $p=300$; (ii) $p=1500$. This is used to
investigate the impact of dimensionality $p$. In model 1b, we set
$(\alpha_1,\alpha_2,\alpha_{10})^\top =(1,1,1)/\sqrt3$ and $\alpha_j=0$ when
$j\neq 1,2,10$.  To evaluate the influence of correlation among covariates,
three scenarios for model 1b are considered.
In scenarios (i) and (ii), $\bX$ and $\bZ$ are generated in the same way
as in model 1a, but now we take $p=1000$, $\sigma_{i j}=\rho^{|i-j|}$, and
consider (i) $\rho=0.25$ and (ii) $\rho=0.75$.
In scenario (iii), $\bX$ and $\bZ$ are correlated.
We generate $\bV=(V_1, V_2, \dots, V_{p+2})^\top $ from normal
distribution with zero mean and covariance matrix
$\boldsymbol{\Sigma}_V=\left(\sigma_{i j}\right)_{(p+2) \times (p+2)}$
with $\sigma_{ij}=0.75^{|i-j|}$.
We let $\bZ=(Z_1,Z_2)^\top =(V_3,V_4)^\top $,
$\bX=(X_1,X_2,X_3\dots,X_p)^\top =(V_1,V_2,V_5,\dots,V_{p+2})^\top $. In real data analysis in
Section 4.2, we use 10-fold cross-validation to choose the bandwidth. But this
is too time-consuming in simulation. As an alternative, we generated several
data sets to get an idea about the range of an appropriate bandwidth. In
simulation study of model 1a, we set the bandwidth $h=0.37$. Note that the standard error
of $\bal^\top \bX$ is 1.08 in this simulation setting. So 95\% values of
$\bal^\top\bX$ lies between $-2.1$ to  2.1. This implies that for a given $t_0$,
we estimate $\eta(t_0)$ based on about 9 percent of observations. Similarly in model 1b, the bandwidth $h$ is set to be $0.37,0.44,0.44$ for each scenario so that
we use about 9 percent of data to estimate $\eta(t_0)$ for a given $t_0$.


\afterpage{
\clearpage
\begin{table}[htbp]
\setlength{\tabcolsep}{1mm}{
\caption{Summary statistics for parameter estimates in model 1a and model 1b.   T: the ratio of the true nonzero coefficients  correctly set to nonzero (\%); F: the ratio of the true zero coefficients incorrectly set to nonzero (\%); Bias ($\times10^{-2}$) and the corresponding mean squared error ($\times10^{-4}$)}\label{est}
\begin{tabular}{cccccccccc}
\toprule
               &           Scenario         & $c_1$  &T&F&$\beta_1$ & $\beta_2$& $\alpha_1$ & $\alpha_2$ & $\alpha_{10}$  \\ \hline
\multirow{9}{*}{$1a$} & \multirow{3}{*}{(i)}
&0&100&0.02&0.34(55)&0.08(57)&-0.01(5.1)&-0.12(14)&-0.27(12)      \\
                    &                            & 0.1 &     100&0.02&-0.24(79)&-0.18(71)&0.13(5.1)&-0.39(16)&-0.33(15)      \\
                    &                            & 0.5 &   100&0.02&0.54(64)&-0.88(56)&-0.14(4.3)&0.06(15)&-0.24(13)\\ \cline{2-10}
                    & \multirow{3}{*}{(ii)}      &0&99&0.05&-0.38(57)&0.01(55)&0.35(10)&-0.80(43)&-1.10(42)   \\
                    &                            & 0.1 &    99&0.01&0.12(55)&0.50(52)&0.16(5.8)&-0.58(30)&-0.42(18)       \\
                    &                            & 0.5 &99&0.01&0.17(59)&0.04(62)&0.17(8.0)&-0.31(32)&-1.10(43)\\ \cline{2-10}
                    & \multirow{3}{*}{Oracle} &0&100&0&0.32(56)&0.04(57)&-0.09(3.9)&-0.06(11)&-0.05(7.6)       \\
                    &                            & 0.1 &100 &0&0.31(61)&0.03(59)&-0.04(4.0)&-0.10(11)&-0.11(8.3)       \\
                    &                            & 0.5 &    100    &   0    &-0.36(55)&-0.17(51)&-0.23(4.6)&0.18(9.2)&-0.01(9.4)      \\ \hline\hline
\multirow{18}{*}{$1b$} & \multirow{3}{*}{(i)}       & 0   &   99&0.09&0.27(17)&-0.1(14)&-0.33(24)&-0.65(32)&-0.88(36)     \\
                    &                            & 0.1 &     98&0.05&0.41(19)&-0.07(17)&-0.43(47)&-0.82(44)&-0.7(39)    \\
                    &                            & 0.5 &    99&0.03&-0.52(31)&-0.51(29)&-0.95(45)&-0.74(46)&-0.97(48)        \\ \cline{2-10}
                    & \multirow{3}{*}{Oracle(i)} & 0   &    100&0&0.17(14)&-0.02(14)&0.07(9.4)&-0.19(9.1)&-0.10(7.2)\\
                    &                            & 0.1 &   100     &    0    &0.18(15)&-0.05(15)&         -0.17(9.2)&-0.13(9.0)&0.08(7.0)      \\
                    &                            & 0.5 &    100    &    0    &-0.16(14)&-0.13(13)&         0.09(11)&-0.07(10)&-0.27(7.3)       \\ \cline{2-10}
                    & \multirow{3}{*}{(ii)}      &  0   &    98&0.06&-0.38(31)&0.17(49)&0.24(42)&0.18(41)&-0.86(38)  \\
                    &                            &  0.1   &   99&0.08&0.19(30)&-0.37(47)&0.25(42)&0.15(41)&-0.85(36)    \\
                    &                            &  0.5   &   98&0.09&-0.28(31)&-0.12(49)&-0.33(41)&0.49(49)&-0.46(36)      \\ \cline{2-10}
                    & \multirow{3}{*}{Oracle(ii)} & 0   &     100   &   0  &0.17(31)&0.01(30)&   0.08(21)&-0.17(22)&-0.34(7.2)      \\
                    &                            & 0.1 &     100   &    0 &0.20(31)&-0.10(32)&   0.20(19)&-0.21(20)&-0.39(6.5)       \\
                    &                            & 0.5 &     100   &    0 &-0.16(29)&-0.03(27)&   0.07(22)&-0.24(51)&-0.52(7.2)       \\ \cline{2-10}
                    & \multirow{3}{*}{(iii)}      &  0  &    98&0.06&0.10(37)&-0.56(30)&-0.41(43)&0.37(44)&-0.73(43)  \\
                    &                            &  0.1  &   96&0.06&0.10(38)&0.32(32)&0.23(43)&0.25(42)&-0.62(32)     \\
                    &                            &  0.5 &  97&0.05&0.01(40)&0.36(29)&0.19(40)&-0.16(41)&-0.35(37)     \\
                    \cline{2-10}
                    & \multirow{3}{*}{Oracle(iii)} & 0   &     100   &   0     &-0.06(44)&-0.36(34)&         -0.03(21)&0.33(22)&-0.74(7.0)       \\
                    &                            & 0.1 &    100   &     0   &0.38(38)&-0.43(34)&         0.10(19)&-0.15(20)&-0.34(6.8)       \\
                    &                            & 0.5 &    100    &    0    &-0.23(38)&-0.09(30)&         0.23(20)&0.01(21)&-0.64(7.2)       \\ \toprule

\end{tabular}}
\end{table}
\thispagestyle{empty}
\clearpage
}

We first examine the finite sample performance of the proposed estimators.
Table \ref{est} reports the ratio of the true nonzero coefficients correctly set to nonzero, denoted
by `T' in the table, the ratio of the true zero coefficients incorrectly set to nonzero, denoted by `F',
bias and mean squared error of the resulting estimates over the 500 replications for
both models under different scenarios and $\bbeta=c_1 \mathbf{1}_2$.
In model 1a, two scenarios share the same oracle model. The ratio of the true nonzero
coefficient that were correctly set to nonzero is always close to 1 and the ratio of
the truly zero coefficients that are incorrectly set to nonzero is always close to 0,
indicating that our method can always correctly identify the true submodel.
Compared with the oracle estimator, both bias and mean squared error of $\hat\bbeta$
behave similar to the oracle one. Bias and mean squared error of $\hat\bal$ is
usually a little bit larger than oracle setting. The error mainly comes from the partial penalization
on $\bal^{(1)}$ using the SCAD penalty and also from the limitation of finite sample and imperfect selection of tuning parameter $\lambda$.
However, the error is in a reasonable order and is acceptable.
We find from the empirical size and power for
(\ref{hyth}) below that the small error does not affect much statistical inference
on parameters of interest.

Further we consider the following null and alternative hypotheses:
\begin{eqnarray}\label{hyth}
H_{01}: \bbeta=0\text{\ \ versus\ \ } H_{11}:\bbeta=c_1 \mathbf{1}_2.
\end{eqnarray}

where $c_1=0,0.01,\cdots, 0.05, 0.1, 0.15,\cdots, 0.5$ for model 1a and $c_1=0,0.01, \cdots, 0.05,\\ 0.1, 0.15, 0.2.$ for model 1b.
When $c_1=0$, it corresponds to $H_{01}$, thus we can examine Type I error rate.
When $c_1\neq 0$, it corresponds to the alternative hypothesis, which allows us
to examine the power of the proposed test.

\begin{figure}[h]
  \centering
  \includegraphics[width=9cm,height=7cm]{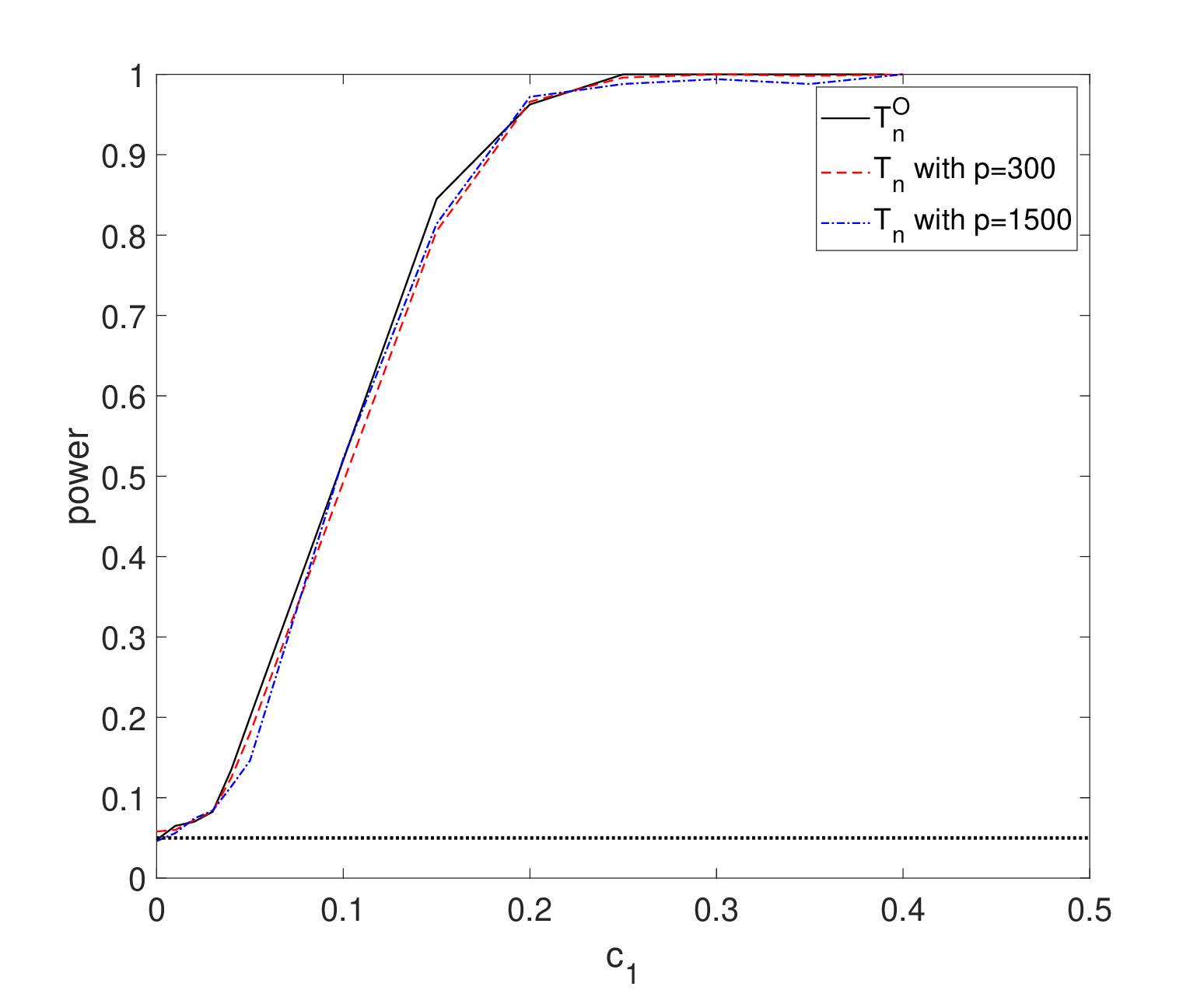}\\
  \caption{\emph{Empirical sizes and powers of $T_n$ and $T^O_n$
  at significance level $0.05$ over 500 replications under model 1a.
  The horizontal dotted line represents level of $0.05$. The solid
line, the long-dashed line, and the dashed-dotted line represent the sizes and
powers of $T^O_n$, $T_n$ with $p=300$ and $T_n$ with $p=1500$, respectively}}\label{1a}
\end{figure}

Figure~\ref{1a} depicts the empirical sizes and powers of
the tests under model 1a. As expected, $T_n$ controls the size well
and is powerful since it performs as well as the oracle test $T^O_n$.
The empirical power of $T_n$  remains roughly the same when $p$ increases from 300 to
1500. This implies that the dimension of $\bX$ does not have a dramatic impact
on the performance of $T_n$.

\begin{table}[htbp]
\setlength{\tabcolsep}{1mm}{
\caption{Empirical sizes and powers of tests $T_n$ and $T^O_n$ under model 1b}\label{tab1b}
\begin{center}
\begin{tabular}{llccccccccc}
\toprule
                      & $c_1$                 & 0     & 0.01  & 0.02  & 0.03  & 0.04  & 0.05  & 0.10  & 0.15 & 0.20 \\ \hline
\multirow{2}{*}{(i)}  & $T^O_n$ & 0.042 & 0.080 & 0.118 & 0.192 & 0.360 & 0.466 & 0.980 & 1    & 1    \\ \cline{2-11}
                      & $T_n$        & 0.058   & 0.086   & 0.109  & 0.266   & 0.343  &  0.506  &  0.937 &   1  &  1        \\ \hline
\multirow{2}{*}{(ii)} & $T^O_n$ & 0.058 & 0.078 & 0.150 & 0.286 & 0.442 & 0.656 & 0.998 & 1    & 1    \\ \cline{2-11}
                      & $T_n$        & 0.042 & 0.068 & 0.116 & 0.196 & 0.406 & 0.604 & 0.994 & 1    & 1    \\ \hline
\multirow{2}{*}{(iii)} & $T^O_n$ & 0.048 & 0.060 & 0.122 & 0.194 & 0.312 & 0.428 & 0.998 & 1    & 1    \\ \cline{2-11}
                      & $T_n$        & 0.052 & 0.076 & 0.125 & 0.188 & 0.296 & 0.436 & 0.960 & 1    & 1    \\ \bottomrule
\end{tabular}
\end{center}
}
\end{table}

Empirical sizes and powers of tests under model 1b  are demonstrated in Table \ref{tab1b}, from which it
can be seen clearly that the empirical sizes of $T_n$ are close to the nominal
level 0.05. Furthermore, the powers of $T_n$ and $T^O_n$ are  close. When
the covariance matrix structure changes with $\rho$ increasing from 0.25 to 0.75, the
powers of $T_n$ and $T^O_n$ increase. This implies that  covariance matrix
of covariates has some impact on  powers of $T_n$ and $T^O_n$. When $\bX$ and $\bZ$ are correlated, there is little power loss. In summary, the proposed test $T_n$ performs as well as the oracle test $T^O_n$ in terms of
power, and controls the empirical sizes very well.

\textbf{Example 2}.  This example is designed to study the performance of the
nonparametric component test statistic $V_n$. To this end, we generate data
from

\textit{Model 2.} $Y= \eta(\boldsymbol{\alpha}^\top \bX)+0.5Z_1-0.3Z_2+0.75\epsilon$.

In this example, $\bX$ and $\bZ$ are generated in the same way as in model 1.a. The dimension
of $\bX$, $p$ is chosen to be 1500 here. We
consider hypotheses:
$$
H_{0} : \eta(t)=\zeta t \quad \text { versus } \quad H_{1} : \eta(t)=c_{2}
\sin \{\pi(t-a) /(b-a)\}+\zeta t,
$$
where $c_2=0,0.01,\cdots,0.05,0.1,0.15,\cdots,0.5$, $a$ and  $b$ are chosen to
be 1.3409 and 0.3912 respectively. This model setting was used by
\cite{liang2010estimation} with fixed $p$. True value of $\zeta$ is chosen to
be 1. Again, when $c_2=0$, the null hypothesis is true, while if $c_2\neq 0$,
the alternative hypothesis holds.


\begin{figure}[h]
  \centering
  \includegraphics[width=9cm,height=7cm]{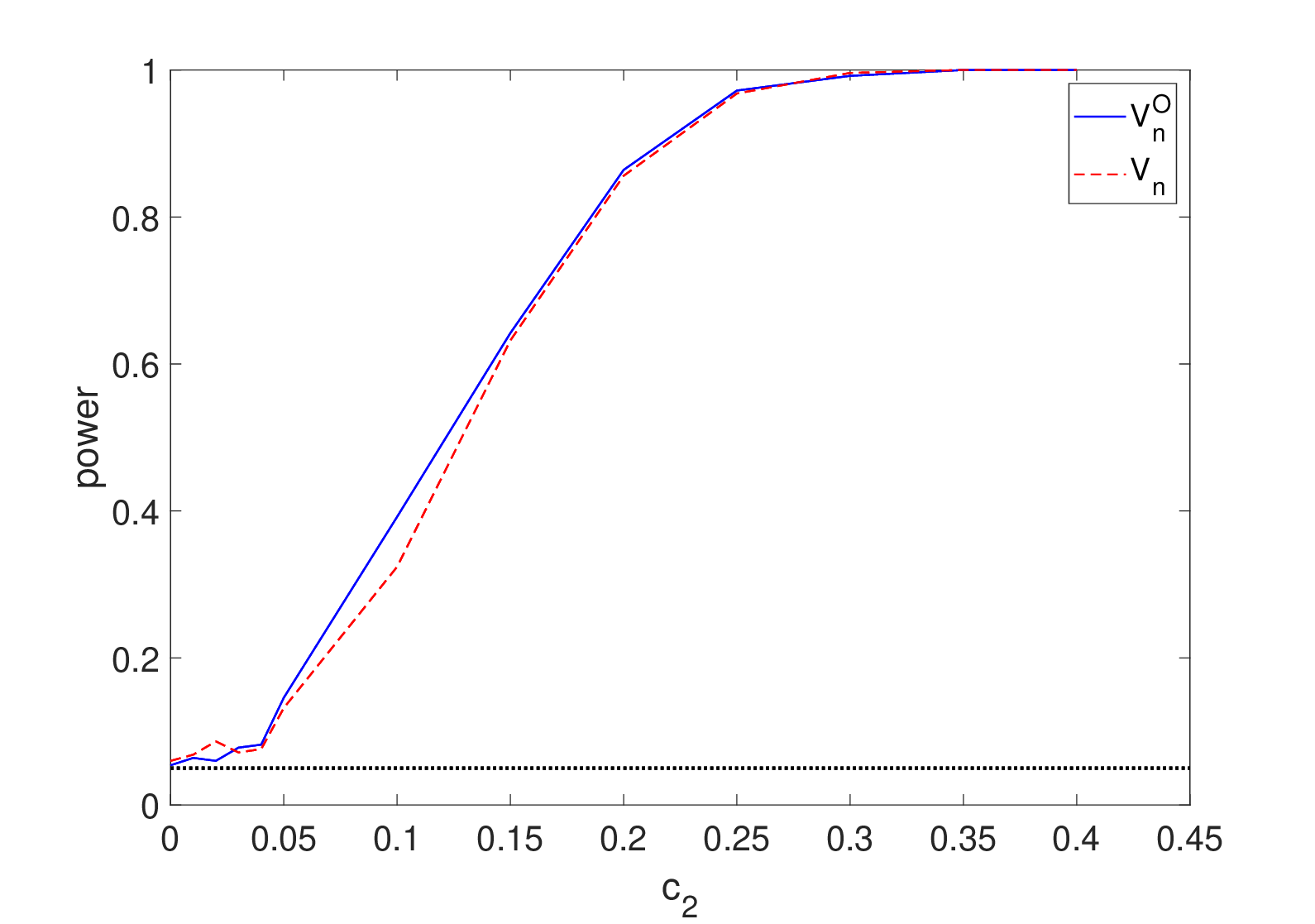}\\
  \caption{\emph{Empirical sizes and powers of $V_n$ and $V^O_n$ at the significance level $0.05$
  over 500 replications under model 2. The horizontal dotted line represents
  the level $\alpha=0.05$. The solid
line and the longdashed line represent the sizes and powers of $V^O_n$ and $V_n$,
respectively}}\label{2a}
\end{figure}

Figure~\ref{2a} depicts the empirical sizes and powers of $V_n$ and $V^O_n$
at significance level 0.05. Figure~\ref{2a} indicates that the
empirical size of $V_n$ is close to the level 0.05. The empirical power of
$V_n$ is greater than 0.95 when $c_2$ is greater than 0.30. Further, the
empirical power of $V_n$ is very close to that of the oracle test $V^O_n$.

\subsection{A real data example}

We now illustrate the proposed methodology by an application to the supermarket data set \citep{wang2009forward}. The data consists of $n=464$ daily records in a supermarket. Each record corresponds to the number of customers as the response and sale amount of 6398 products as predictors. We standardize both response and predictors to be with zero mean and unit variance. We fit a partially linear single-index model to the data set and aim to locate  products whose sale volumes are mostly
correlated with the number of customers, and perform related hypothesis testings to the regression function.

\begin{figure}[t]
  \centering
  \includegraphics[width=8cm,height=7cm]{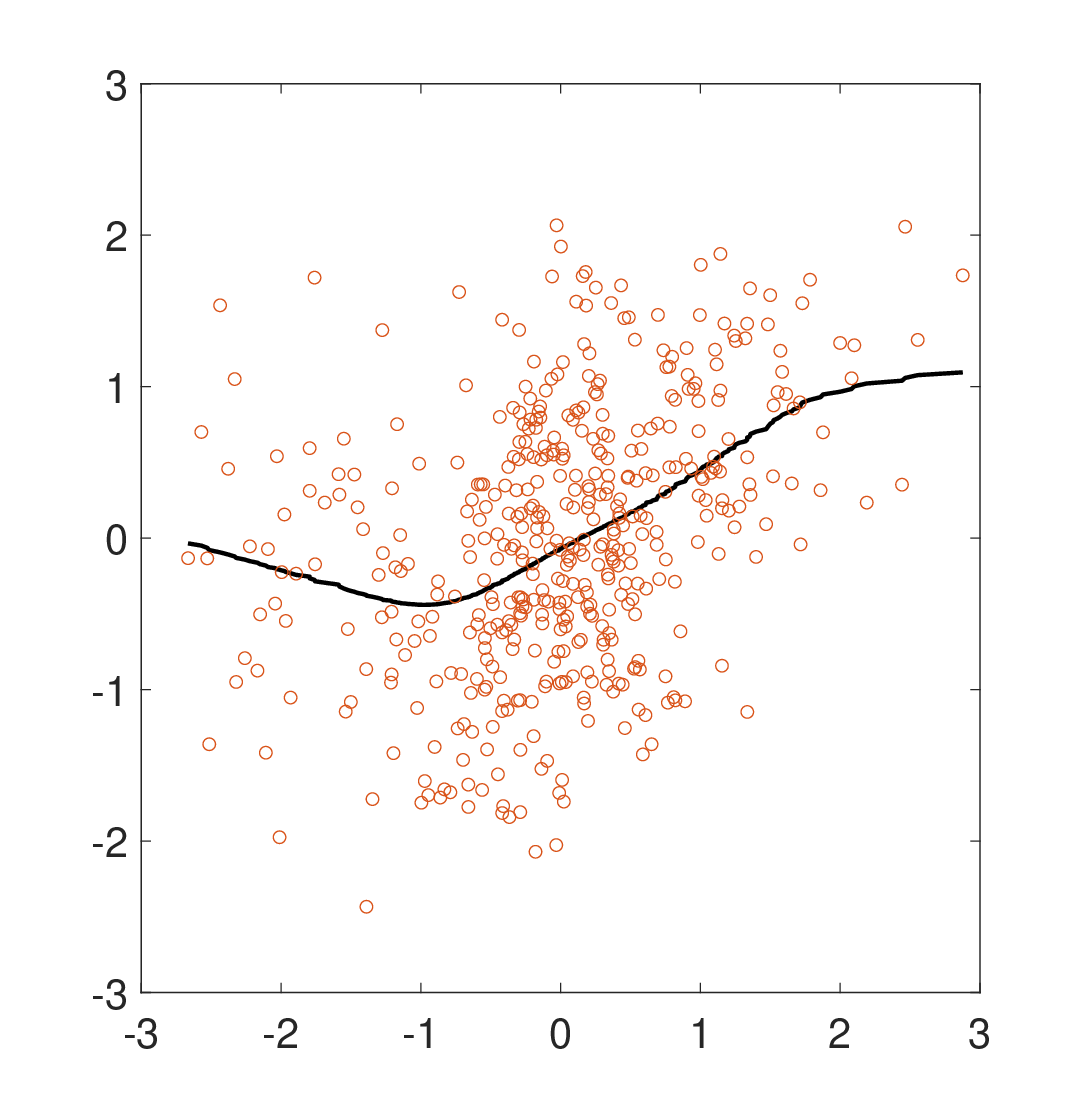}\\
  \caption{\emph{Scatter plot of $\hat{\bal}^\top\bX$ versus $Y-\hat\bbeta^\top\bZ$. Solid line is the estimate $\hat\eta(\hat{\bal}^\top\bX,\hat{\boldsymbol{\theta}})$ of $\eta({\bal}^\top\bX)$}}\label{F3}
\end{figure}

We carry out a preprocessing step that we reduce the dimension of predictors to 1000 by employing the feature screening scheme in \citet{li2012feature}. To decide which variables belong to the linear part $\bbeta^\top \bZ$ or the
nonlinear part $\eta(\bal^\top \bX)$, we adopt the strategy suggested by
\cite{xia1999extended} and \cite{zhang2012dimension}. Their ideas are based on
dealing with the scatterplot of each variable versus the response. To
take both linear behavior and goodness of fit into account, we first compute
the Pearson correlation between the response $Y$ and each variable and select
variables with an absolute value of the correlation greater than 0.3. Then we fit
the response $Y$ and each covariate with local linear smoothing. We obtain the
estimated mean function and corresponding pointwise confidence band by
computing the mean plus and minus $k$ times the estimated standard deviation
function. We also fit a linear regression. If the linear straight line lies in
the confidence band, the variable goes into the linear part of our model. In this
example, we choose $k= 0.3$ as the threshold. Otherwise, if the correlation is
smaller than 0.3 or the linear regression line cannot be encompassed by the
confidence band, the variable goes into the single-index part. This leads to a
number of 994 variables ($\bX$) for single-index component and 6 variables
($\bZ$) for the linear component. Model 3 below is fitted.
We use 10-fold cross-validation to choose the bandwidth and HBIC to choose $\lambda$. This leads to the selected bandwidth $h=0.59$ and the selected tuning parameter
$\lambda=0.12$. The fit of the semiparametric
part of the model is shown in Figure \ref{F3}. There are in total of 12 active variables selected from covariates $\bX$.

\textit{Model 3.}
    $Y=\eta({\alpha}^\top\bX)+\boldsymbol{\beta}^\top\bZ+\epsilon.$

From Figure \ref{F3}, there seems to exist a nonlinear pattern for the
relationship between $\bX$ and the response $Y$.

We apply the proposed hypothesis testing procedures for this data set. We
first consider the following two testing problems: $H_{01}: \bbeta=0$ and
$H_{02}: \eta (t)=c$, where $c$ is a constant. These two hypotheses $H_{01}$
and $H_{02}$ aim to test whether the covariates $\bZ$ and $\bX$ contribute to the
response $Y$ or not, respectively. Based on the proposed test statistics $T_n$
and $V_n$, the $p$-values of the two hypotheses are 0. This implies that the
selected variables indeed show a significant influence on the response. Then we
further test whether the contribution of the single-index components
$\bX$ to the response $Y$ is linear or not and perform hypothesis testing
$H_{03}:\eta (t)=\zeta t$. The corresponding $p$-value is 0.004.
Therefore, we reject the null hypothesis that the link function for single-index
part is linear at level 0.05.  This is consistent with our observation on the
fitting plot of the semiparametric part shown in Figure~\ref{F3}. In conclusion, we find there
is a significant effect of the selected variables and potential non-linear
relationship between selected variables and the response. It is worth pointing out that \citet{liu2016global} and \citet{lan2016testing} did the analysis on the same data set but fitted high dimensional linear model on it. However, from our analysis, we find evidence that the linear model is not suitable and recommend fitting a semiparametric model on this data and data with similar attributes.

\section{Conclusions and discussions}

In this paper, we developed new statistical inference procedures for high
dimensional partially linear single-index model. Different from  linear
regression model, we have to deal with the nuisance unknown function. To derive
powerful hypothesis tests, we first propose a profile penalized least squares
estimator and study its asymptotic properties. Then we propose an F-type test
for the parametric component. We further study the specification testing
problem for the nonparametric component and propose a test statistic with an
asymptotic normal distribution. A notable feature is that the optimal rate for
the bandwidth is allowed, even the covariate $\bX$ is of high dimensional. No
under-smoothing or over-smoothing is required. The dimensionality is allowed to be exponential order of the sample size and the sparsity level can also be diverging.

In this paper, we focus on testing the low
dimensional parametric vector $\bbeta$,  while regard the high
dimensional parametric vector $\bal$ as nuisance parameter. In
practice, it may also be interesting to make inference on
components of $\bal$. This issue was studied recently by
\cite{eftekhari2021inference} for single index model under
elliptical symmetry assumption on the covariates. The extension of
their procedure to PLSIM without elliptical assumption would be an
interesting topic for future research.



\section*{Supplement}
The supplementary material consists of three technical lemmas, the proofs of these three lemmas and proofs of all theorems in this paper.

\section*{Appendix: Technical Proofs}

\def\thelemma{A.\arabic{lemma}}
\def\thepro{A.\arabic{pro}}
\renewcommand{\thesection}{A}
\renewcommand{\theequation}{A.\arabic{equation}}
\renewcommand{\thetable}{A\arabic{table}}
\renewcommand{\thefigure}{A\arabic{figure}}

\setcounter{figure}{0}
\setcounter{table}{0}

The following three lemmas will be used in the proofs for the main theorems, and their
proofs are given in the supplementary material of this paper.

We first establish some results on uniform convergence for kernel estimation,
which has been considered in several authors \citep{mack1982weak,
liebscher1996strong, hansen2008uniform}, but none of them considered estimating several regression
functions simultaneously with the number of regression functions growing with $n$.
Thus, their results cannot be directly applied for nonparametric estimation in the presence of
high dimensional covariates. In Lemma \ref{Lemma0} below, we establish
the uniform convergence rate for kernel estimation for the regression functions when the number
of regression functions grows with sample size $n$. Let $X_0$ and $\bY_0$ be
a $1$- and $s$-dimensional continuous random vector, respectively.
With a slight abuse of notation, here $X_0$ and $\bY_0$ represent general random variable and vector, respectively, and
are not the covariate and the response  in the main text.

\begin{lemma}\label{Lemma0}
Suppose that $\{\{X_i,\bY_i\}, i = 1,\cdots, n\}$ is a random sample from $\{X_0,\bY_0\}$,
where the dimension of $\bY_0=(Y_{01},\cdots, Y_{0s})^\top$  grows with $n$.
Let $K(\cdot)$ be a kernel function, and
\[
\bH(x)= \dfrac{1}{nh} \sum_{i=1}^{n}  K\left(\dfrac{X_{i}-x}{h}\right)\bY_{i}.
\]
It follows that
\begin{eqnarray}\label{sdot5}
\sup _{x}\|\bH(x)-E\bH(x)\|_2=O_{p}\left(\left\{\frac{s\log n}{n h}\right\}^{1 / 2}\right)
\end{eqnarray}
under the following three assumptions:
\begin{description}
\item {\bf Assumption 1}. The density of $X_0$, $f(x)$, satisfies that $\sup_x f(x)\leq B_0<\infty$.
For some $M>2$ and $q>0,\delta>0,$ $E|X_0|^{2q+\delta}<\infty$, and
\begin{eqnarray*}
&&\sup_{1\le k\le s}\sup_xE(|Y_{0k}|^M|X_0=x)f(x)\leq B_1<\infty,\\
&&\sup_{1\le k\le s}\sup_x|x|^qE(Y_{0k}|X_0=x)f(x)\leq B_2<\infty.
\end{eqnarray*}
\item {\bf Assumption 2}.
$K(u)$ is differentiable and $\int|K^2(u)|<\infty$. $|K'(u)|\leq \lambda_1<\infty$. There exist
some constants $\eta>1$, $L>0$ and $\lambda_2<\infty$ such that
$|K'(u)|\leq \lambda_1|u|^{-\eta}$, $|K(u)|\leq \lambda_2|u|^{-q}\leq\bar{K}<\infty$ for
 $|u|>L$.
 \item {\bf Assumption 3}.
 The bandwidth satisfies $h\rightarrow 0$ and  $s(\log n)^{1+\delta}/(n^{1-2/M} h)\rightarrow 0$ for some $\delta>0$.

\end{description}

\end{lemma}
We introduce some notation for the following. Let $\Gamma_{i0}=\bal_0^\top\bX_i$, $\Gamma_{i1}=\bal^\top\bX_i, \eta_{i0}=\eta(\Gamma_{i0}), \eta_{i1}=\eta(\Gamma_{i1}), \hat\eta_{i0}=\hat\eta(\Gamma_{i0},\btheta_0)$ and $\hat\eta_{i1}=\hat\eta(\Gamma_{i1},\btheta)$. Further let
$W_{0ij}=W_{nj}(\Gamma_{i0};\bal_0)$ and $W_{1ij}=W_{nj}(\Gamma_{i1};\bal)$.  Denote $\mu_{1}(\Gamma_{i1})=E[\bX_{i,\A^*}|\Gamma_{i1}], \widetilde \bX_i(\bal)=
\bX_{i,\A^*}-\mu_{1}(\Gamma_{i1}), \mu_{2i}(\Gamma_{i1})=E[\bZ_i|\Gamma_{i1}]$,
$\widetilde \bZ_i(\bal)=\bZ_i-\mu_{2}(\Gamma_{i1})$, $\mu^{\star}_{1i}=\mu_1(\Gamma_{i0}),
\mu^{\star}_{2i}=\mu_2(\Gamma_{i0}), \widetilde \bX^{\star}_i=\widetilde \bX_i(\bal_0),$
$\widetilde \bZ^{\star}_i=\widetilde \bZ_i(\bal_0)$. Define
\begin{eqnarray*}
L_i=\left(
                         \begin{array}{c}
                           \widetilde \bZ_i^{\star}\\
                           \eta'_{i0}J^\top _{0,\mathcal{A}}\widetilde \bX_i^{\star}\\
                         \end{array}
                       \right);\,\,\,
\hat L_{i}=\left(
  \begin{array}{ccc}
    \bZ_i+\dfrac{\partial \hat\eta_{i1}}{\partial\bbeta}\\
   \dfrac{\partial \hat\eta_{i1}}{\partial\bal^{(1)}_{\mathcal{A}}}\\
  \end{array}
\right).
\end{eqnarray*}
\begin{lemma}\label{Lemma1}
Under conditions (A1), (A4) and (A5), for any $\btheta$ which satisfies
$\bal^{(1)}_{\A^c}=0$ and $\|\bvarth-\bvarth_0\|_2=O(\sqrt{s/n})$, it follows that
\begin{eqnarray*}
(\hat \eta_{i0}-\eta_{i0})^2=O_p(h^4+\frac{\log n}{nh});\,\,\,\|\hat
L_i-L_i\|^2_2=O_p(s(h^4+\frac{\log n}{{nh^3}}))
\end{eqnarray*}
uniformly for $i$.
\end{lemma}

\begin{lemma} \label{Lemma2}
Under conditions (A1), (A4), (B2) and (B3), for any $\btheta$ which satisfies
$\bal_{\A^c}=0$ and $\|\btheta-\btheta_0\|_2=O(\sqrt{s/n})$, we have:
\begin{eqnarray*}
\|\hat\zeta-\zeta_0\|_2=O_p(\sqrt{\frac{s}{n}}).
\end{eqnarray*}
\end{lemma}

\vspace{2mm}

{\noindent\bf{Proof of Theorem 1}:}

To enhance the readability, we divide the proof of Theorem 1 into three steps. In the first step,
we show that there exists a local
minimizer $\bar{\btheta}$ of $Q_n(\btheta)$ with the constraints $\bar{\bal}^{(1)}_{\mathcal{A}^c}=0$,
such that $\|\bar{\btheta}-\btheta_0\|_2=O_p(\sqrt{s/n})$. In the second step, we prove that
$\bar{\btheta}$ is indeed a local minimizer of $Q_n(\btheta)$. This implies $\hat\btheta=\bar{\btheta}$. In the final step, we derive the asymptotic expansion of $\hat\btheta$.

\vspace{3mm}

\emph{Step 1: Consistency in the $(s+q)$-dimensional subspace:} We first
constrain $Q_n(\btheta)$ on the $(s+q)$-dimensional subspace
of $\{\btheta\in\mathbb{R}^{p+q-1}:\bal^{(1)}_{\mathcal{A}^c}=0\}$ of $\mathbb{R}^{p+q-1}$. This partial penalized least squares function is given by
$$\bar{Q}_n(\bvarth)=\frac{1}{2n}\sum_{i=1}^n [Y_i-\hat\eta(\bal^\top_{\A^*}\bX_{i,\A^*},\bvarth)-\bbeta^\top \bZ_i]^2+\sum_{j=1}^s p_{\lambda}(|\delta_j|).$$
Here $\bvarth=(\bbeta^\top ,\bdelta^\top )^\top $ and $\bdelta=(\delta_1,\cdots,\delta_s)^\top $. We now show that there exists a strict local
minimizer $\bar{\bvarth}$ of $\bar{Q}_n(\bvarth)$ such that $\|\bar\bvarth-\bvarth_0\|_2=O_p(\sqrt{s/n})$.
To this end, we consider an event
$$H_n=\{\min_{\bvarth\in\partial\mathcal{N}_{\tau}}\bar{Q}_n(\bvarth)>\bar{Q}_n(\bvarth_0)\}.$$
where $\mathcal{N}_{\tau}=\{\bvarth\in\mathbb{R}^{s+q}: \bvarth=\bvarth_0+d_n\bm{u}, \|\bm{u}\|_2\leq \tau\}$ with $d_n=\sqrt{s/n}$, $\tau\in (0,\infty)$, and
$\partial\mathcal{N}_{\tau}$ denotes the boundary of the closed set $\mathcal{N}_{\tau}$. Clearly, on the event $H_n$, there exists a local minimizer of $\bar{Q}_n(\bvarth)$ in $\mathcal{N}_{\tau}$.
Thus, we only need to show that $P(H_n)\rightarrow 1$ as $n\rightarrow \infty$ when $\tau$ is large.
To this end, we next study the behavior of
$\bar{Q}_n$ on the boundary $\partial\mathcal{N}_{\tau}$.

Define
\begin{eqnarray}\label{eqnA.1}
D_{n}=\frac{1}{n}\sum_{i=1}^n[Y_i-\hat\eta_{i1}-\bbeta^\top \bZ_i]^2-\frac{1}{n}\sum_{i=1}^n[Y_i-\hat\eta_{i0}-\bbeta^\top _0\bZ_i]^2.
\end{eqnarray}
Note that
\begin{eqnarray}\label{eqnA.2}
D_{n}&=&\frac{1}{n}\sum_{i=1}^n[\hat\eta_{i1}-\hat\eta_{i0}+(\bbeta-\bbeta_0)^\top \bZ_i]^2\nonumber-\frac{2}{n}\sum_{i=1}^n[\hat\eta_{i1}-\hat\eta_{i0}+(\bbeta-\bbeta_0)^\top \bZ_i][Y_i-\hat\eta_{i0}-\bbeta^\top _0\bZ_i]\nonumber\\
&=:&D_{n1}-2D_{n2}.
\end{eqnarray}
Let $\bal^*$ be between $\bal$ and $\bal_0$, $\Gamma^*_i=\bal^{*\top}\bX_i,
\eta^*_{i}=\eta(\Gamma^*_i), \hat\eta^*_{i}=\hat\eta(\Gamma^*_i,\btheta^*)$,
and
\begin{eqnarray*}
L^*_{i}=\left(
  \begin{array}{ccc}
    \bZ_i+\dfrac{\partial \hat\eta^*_{i}}{\partial\bbeta}\\
   \dfrac{\partial \hat\eta^*_{i}}{\partial\bal^{(1)}_{\mathcal{A}}}\\
  \end{array}
\right).
\end{eqnarray*}
We have that
\begin{eqnarray*}
&&\hat\eta_{i1}-\hat\eta_{i0}+(\bbeta-\bbeta_0)^\top \bZ_i=(\bvarth-\bvarth_0)^\top L_i+(\bvarth-\bvarth_0)^\top (L^*_i-L_i);\\
&&Y_i-\hat\eta_{i0}-\bbeta^\top _0\bZ_i=\epsilon_i+\eta_{i0}-\hat\eta_{i0}.
\end{eqnarray*}
Thus it follows that
\begin{eqnarray}
D_{n1}&=&(\bvarth-\bvarth_0)^\top \frac{1}{n}\sum_{i=1}^nL_iL_i^\top (\bvarth-\bvarth_0)\nonumber + 2(\bvarth-\bvarth_0)^\top \frac{1}{n}\sum_{i=1}^nL_i(L^*_i-L_i)^\top (\bvarth-\bvarth_0)
\nonumber\\
&& +(\bvarth-\bvarth_0)^\top \frac{1}{n}\sum_{i=1}^n(L^*_i-L_i)(L^*_i-L_i)^\top (\bvarth-\bvarth_0)\\
&=:&D_{n11}+ D_{n12} + D_{n13}, \label{eqnA.3}
\end{eqnarray}
and
\begin{eqnarray}
 D_{n2} &=&(\bvarth-\bvarth_0)^\top \frac{1}{n}\sum_{i=1}^nL_i\epsilon_i+
(\bvarth-\bvarth_0)^\top \frac{1}{n}\sum_{i=1}^n(L^*_i-L_i)\epsilon_i\nonumber\\
&& +(\bvarth-\bvarth_0)^\top
\frac{1}{n}\sum_{i=1}^nL_i(\eta_{i0}-\hat\eta_{i0})\label{eqnA.4}\\
&&+(\bvarth-\bvarth_0)^\top  \frac{1}{n}\sum_{i=1}^n(L^*_i-L_i)(\eta_{i0}-\hat\eta_{i0})\\
 &=:&D_{n21}+D_{n22}+D_{n23}+D_{n24}. \nonumber
\end{eqnarray}

In what follows, we will show that $D_{n12}, D_{n13}, D_{n22}, D_{n23}$,
and $D_{n24}$ are all of the order $o_p(s/n)$. Thus they are dominated by $D_{n11}$ and $D_{n21}$.

It follows by the Cauchy-Schwarz inequality that
\begin{eqnarray}\label{eqnA.5}
|\frac{1}{2}D_{n12}|^2&\leq& \frac{1}{n}\sum_{i=1}^n[\{(\bvarth-\bvarth_0)^\top L_i\}^2]\frac{1}{n}\sum_{i=1}^n\{(\bvarth-\bvarth_0)^\top (L^*_i-L_i)\}^2\nonumber\\
&\leq&\frac{1}{n}\sum_{i=1}^n[\{(\bvarth-\bvarth_0)^\top L_i\}^2]\frac{1}{n}\sum_{i=1}^n||L_i^*-L_i||_2||\bvarth-\bvarth_0||_2
\end{eqnarray}
From Lemma~\ref{Lemma1} and condition (A1), it follows that
\begin{eqnarray*}
&&E[(\bvarth-\bvarth_0)^\top L_i]^2=(\bvarth-\bvarth_0)^\top E(L_iL_i^\top )(\bvarth-\bvarth_0)
=(\bvarth-\bvarth_0)^\top \Sigma^{\star}(\bvarth-\bvarth_0)=O(\frac{s}{n});\\
&&\{(\bvarth-\bvarth_0)^\top (L^*_i-L_i)\}^2\leq \|\bvarth-\bvarth_0\|^2\|L^*_i-L_i\|^2_2=o_p(\frac{s}{n}),
\end{eqnarray*}
hold uniformly for $i$ when $s=o(\sqrt{n})$, $nh^3/(s\log n)\rightarrow \infty$, and $nh^8 \rightarrow 0$.
Thus $D_{n12}=o_p(s/n)$. Similarly we can show that $D_{n13}=o_p(s/n)$. We next
deal with $D_{n24}$. By
\begin{align}\label{eqnA.6}
&D_{n24}^2\leq \frac{1}{n^2}\sum_{i=1}^n\{(\bvarth-\bvarth_0)^\top (L_i^*-L_i)\}^2\sum_{i=1}^n(\eta_{i0}-\hat{\eta}_{i0})^2\\
&\leq||\bvarth-\bvarth_0||_2^2\dfrac{1}{n}\sum_{i=1}^n||L_i^*-L_i||_2^2\dfrac{1}{n}\sum_{i=1}^n(\eta_{i0}-\hat{\eta}_{i0})^2\nonumber\\
&=O_p(\frac{s}{n})O_p(s(h^4+\frac{\log n}{nh^3}))O_p(h^4+\frac{\log n}{nh}))=o_p(\frac{s^2}{n^2}), \nonumber
\end{align}
under condition that $nh^8\rightarrow 0$, $\log n =o(1/h)$ and $(\log n)^2=o(nh^4)$.

The orders of $D_{n22}$ and $D_{n23}$ can be derived using the same argument.
We only show the proof for $D_{n23}$. For the term $D_{n23}$, it follows that
\begin{eqnarray*}
||\frac{1}{n}\sum_{i=1}^nL_i(\eta_{i0}-\hat\eta_{i0})\}||^2_2&=&\frac{1}{n^2}\sum_{i=1}^n(\eta_{i0}-\hat\eta_{i0})^2L_i^{\top}L_i+\frac{1}{n^2}\sum_{i=1}^n(\eta_{i0}-\hat\eta_{i0})(\eta_{j0}-\hat\eta_{j0})L_i^{\top}L_j\\
&=:&D_{n231}+D_{n232}.
\end{eqnarray*}
For the term $D_{n231}$, by Lemma \ref{Lemma1} and condition (A3), \[\frac{1}{n^2}\sum_{i=1}^n(\eta_{i0}-\hat\eta_{i0})^2L_i^{\top}L_i\leq \sup_{1\le i\le n}(\eta_{i0}-\hat\eta_{i0})^2\frac{1}{n^2}\sum_{i=1}^nL_i^{\top}L_i=o_p(s/n).\]
For the term $D_{n232}$, noticing that $E[L_i|\Gamma_{i0}]=0$, and $E[L_i^{T}L_i|\Gamma_{i0}=t]/s$
is bounded uniformly of $t$, we can show $D_{n232}=o_p(s/n)$ applying martingale central limit theorem (Corrollary 3.1  in \citep{hall2014martingale}),

Up to now, we show that $D_{n12},
D_{n13}, D_{n22}, D_{n23}$, and $D_{n24}$ are all of the order $o_p(s/n)$. As a
result, it follows that
\begin{equation}\label{eqnA.7}
D_{n}=(\bvarth-\bvarth_0)^\top \frac{1}{n}\sum_{i=1}^nL_iL_i^\top (\bvarth-\bvarth_0)-2(\bvarth-\bvarth_0)^\top \frac{1}{n}\sum_{i=1}^nL_i\epsilon_i
+o_p(\frac{s}{n}).
\end{equation}
Under conditions $s=o(n^{1/2})$ and $\lambda_{\max}(\Sigma^{\star})<\infty$, we have
$$\|\frac{1}{n}\sum_{i=1}^nL_iL_i^\top -\Sigma^{\star}\|_2=O_p(\frac{s}{\sqrt n})=o_p(1).$$
Further note that
$$\|\frac{1}{n}\sum_{i=1}^nL_i\epsilon_i\|_2=O_p(\sqrt{\frac{s}{n}}).$$

On the boundary $\partial\mathcal{N}_{\tau}$, $\bvarth-\bvarth_0=d_n\bm{u}$, $\|\bm{u}\|_2=\tau$, and thus
\begin{eqnarray}\label{eqnA.8}
D_{n}&=&d^2_n\bm{u}^\top \Sigma^{\star}\bm{u}-O_p(d^2_n)\|\bm{u}\|_2+o_p(\frac{s}{n}).
\end{eqnarray}
In summary, by allowing $\|\bm{u}\|_2=\tau$ to be large enough, all terms of
$D_n$ is dominated by the first term which is positive under condition (A1).

Using Taylor's expansion, we have
\begin{eqnarray*}
&&\sum_{j=1}^s p_{\lambda}(|\delta_j|)-\sum_{j=1}^s
p_{\lambda}(|\alpha^{(1)}_{0j,\mathcal{A}}|)\\
&=&(\bdelta-\bal^{(1)}_{0,\mathcal{A}})^\top
\lambda_n\bar{\rho}(\bal^{(1)}_{0,\mathcal{A}})
+\frac{1}{2}(\bdelta-\bal^{(1)}_{0,\mathcal{A}})^\top
\bm{\Lambda}^*(\bdelta-\bal^{(1)}_{0,\mathcal{A}}).
\end{eqnarray*}
Here $\bm{\Lambda}^*$ is a diagonal matrix.
By condition (A2), the maximum eigenvalue of $\bm{\Lambda}^*$ is bounded
by $\lambda_n\kappa_0=o(1)$.
It follows from the concavity of $\rho(\cdot)$, $l_n<|\alpha^{(1)}_{0j,\mathcal{A}}|$,
and condition (A2) that
\begin{eqnarray*}
\|\lambda_n\bar{\rho}(\bal^{(1)}_{0,\mathcal{A}})\|^2_2\leq (s^{1/2}p'_{\lambda}(l_n))^2=o(\frac{1}{n}).
\end{eqnarray*}
These results imply that \begin{eqnarray*}
\sum_{j=1}^s p_{\lambda}(|\delta_j|)-\sum_{j=1}^s p_{\lambda}(|\alpha^{(1)}_{0j,\mathcal{A}}|)
=o_p(d^2_n).
\end{eqnarray*}

Finally, by allowing $\|\bm{u}\|=\tau$ to be large enough,
we conclude that $\bar{Q}_n(\bvarth)-\bar{Q}_n(\bvarth_0)$ is dominated by a positive value. Consequently, step 1 is obtained.

\vspace{3mm}

\vspace{2mm}
\emph{Step 2: Sparsity.}
According to Theorem 1 in \cite{fan2011nonconcave}, it suffices to show that with probability tending to 1,
we have
\begin{eqnarray}\label{eqnA.9}
\max_{k\in\A^c}|B_{nk}|=:\max_{k\in\A^c}|\frac{1}{n}\sum_{i=1}^n (Y_i-\hat\eta_{i1}-\bbeta^\top\bZ_i)\frac{\partial\hat\eta_{i1}}{\partial\alpha_k}|\ll\lambda_n.
\end{eqnarray}
Here ${\btheta}=({\bbeta}^\top, {\bal}^\top)^\top$ satisfies that ${\bal}^{(1)}_{\A^c}=0$
and $\|{\btheta}-\btheta_0\|_2=O_p(\sqrt{s/n})$.

%


Firstly, define
\begin{eqnarray}\label{eqnA.11}
R_{ki}=\Big(J^\top(\bX_i-E[\bX_i|\bal^\top\bX_i])\Big)_k\eta'_{i1}.
\end{eqnarray}

Secondly note that
\begin{eqnarray*}
&&Y_i-\hat\eta_{i1}-\bbeta^\top\bZ_i=\epsilon_{i}+\eta_{i0}-\eta_{i1}+(\bbeta_0-\bbeta)^\top\bZ_i+\eta_{i1}-\hat\eta_{i1}\\
&=&\epsilon_i+(\eta_{i1}-\sum_{j=1}^n W_{1ij}(\eta_{j1}+\epsilon_j))+\bR^\top_{i,\A}(\bal^{(1)}_{0,\A}-\bal^{(1)}_{\A})+\tilde \bZ^\top_{i}(\bal)(\bbeta_0-\bbeta)
+o_p(\sqrt{\frac{s}{n}}).
\end{eqnarray*}


As a result, we obtain that
\begin{eqnarray*}
B_{nk}&=&\frac{1}{n}\sum_{i=1}^n \epsilon_iR_{ki}+\frac{1}{n}\sum_{i=1}^n [\tilde \bZ^\top_{i}(\bal), \bR^\top_{i,\A}]R_{ki}(\bvarth_0-\bvarth)\\
      &&+\frac{1}{n}\sum_{i=1}^n [\eta_{i1}-\sum_{j=1}^n W_{1ij}(\eta_{j1}+\epsilon_j)]R_{ki}
+\frac{1}{n}\sum_{i=1}^n \epsilon_i[\frac{\partial\hat\eta_{i1}} {\partial\alpha_{k}}-R_{ki}]\\
&&+\frac{1}{n}\sum_{i=1}^n [(\eta_{i0}-\hat\eta_{i1})+(\bbeta_0-\bbeta)^\top \bZ_i][\frac{\partial\hat\eta_{i1}}{\partial\alpha_{k}}-R_{ki}]
+o_p(\sqrt{\frac{s}{n}})\\
&&=:\sum_{i=1}^5 B_{nik}+o_p(\sqrt{\frac{s}{n}}).
\end{eqnarray*}
In the following, we aim to determine the orders of $B_{n1i},i=1,\cdots, 5$.
Let $a_n=n^{1/\varpi+\zeta}$. First, by condition (A3) and Markov inequality, we can show that
$$P(\max_{1\leq i\leq n}\max_{k\in\A^c}|\epsilon_iR_{ki}|>a_n)\leq nE[(\max_{k\in\A^c}|\epsilon_iR_{ki}|)^{\varpi}]/a^{\varpi}_n\rightarrow 0.$$
Let $x=\sqrt{C\log p/n}$ with $C$ being large enough. Then by using Bernstein inequality, we obtain that
\begin{align*}
&P\left(|\sum_{i=1}^n \epsilon_iR_{ki}|>nx,\,\,\mbox{for some}\,\,k\in\A^c\right)\\
&\leq (p-s)\max_kP(|\sum_{i=1}^n \epsilon_iR_{ki}|>nx)\\
&\leq 2p\max_k\exp\left\{-\frac{n^2x^2}{2\sum_{i=1}^n E(R^2_{ki}\epsilon^2_{i})+2nxa_n/3}\right\}\rightarrow 0.
\end{align*}
Thus we get $$\max_k|B_{n1k}|=O_p(\sqrt{\frac{\log p}{n}}).$$
Further we have that
\begin{eqnarray*}
\max_k|B_{n2k}|\leq n^{-1}\|\bR^{\top}_{\mathcal{A}^c}(\tilde{\bZ}(\bal), \bR_{\mathcal{A}})\|_{2,\infty}\|\bvarth_0-\bvarth\|_2
=O_p(a_n\sqrt{\frac{s}{n}}).
\end{eqnarray*}

Now we turn to consider the third term $B_{n3k}$. According to the proof of
Lemma \ref{Lemma1}, $
\eta_{i1}-\sum_{j=1}^n W_{1ij}(\eta_{j1}+\epsilon_j)=O_p(h^2+\sqrt{\dfrac{\log n}{nh}}).
$ holds uniformly over $i$.
Thus, we have $$\max_k|B_{n3k}|=O_p(h^2+\sqrt{\frac{\log n}{nh}}).$$
%

Similarly, we can show that $$\max_k|B_{n4k}|=O_p(a_n(h^2+\sqrt{\frac{\log n}{nh^3}})).$$
Lastly,
$$
\max_k|B_{n5k}|=O_p\left\{(\sqrt{\frac{s}{n}}+h^2+\sqrt{\frac{\log n}{nh}})(h^2+\sqrt{\frac{\log n}{nh^3}})\right\}
=o_p(\sqrt{\frac{s}{n}}),
$$
under condition (A6a). Thus step 2 is finished.

%
\vspace{3mm}

\emph{Step 3: Asymptotic expansions.}
Steps 1 and 2 show that $\hat\bal^{(1)}_{\mathcal{A}^c}=0$ with probability 1, and
further $\|\hat\bal^{(1)}_{\mathcal{A}}-\bal^{(1)}_{0,\mathcal{A}}\|_2=O_p(\sqrt{s/n})$.

First let
\begin{eqnarray}\label{eqnA.14}
\dot{L}(\bvarth_0)=\frac{1}{\sqrt n}\sum_{i=1}^n (Y_i-\hat\eta_{i0}-\bbeta^\top _0 \bZ_i)
\left(
  \begin{array}{ccc}
    \bZ_i+\dfrac{\partial \hat\eta_{i0}}{\partial\bbeta}\\
   \dfrac{\partial \hat\eta_{i0}}{\partial\bal^{(1)}_{\mathcal{A}}}\\
  \end{array}
\right).
\end{eqnarray}
For $\hat\bvarth$,  let
\begin{equation}\label{eqnA.15}
\dot{L}(\hat\bvarth)=\dfrac{1}{\sqrt n}\sum_{i=1}^n (Y_i-\hat\eta_{i1}-\hat\bbeta^\top  \bZ_i)
\left(
  \begin{array}{ccc}
    \bZ_i+\dfrac{\partial \hat\eta_{i1}}{\partial\bbeta}\\
   \dfrac{\partial \hat\eta_{i1}}{\partial\bal^{(1)}_{\mathcal{A}}}\\
  \end{array}
\right)=\left(
  \begin{array}{ccc}
    0\\
   \sqrt n\lambda_n\bar{\rho}(\hat\bal^{(1)}_{\mathcal{A}})\\
  \end{array}
\right).
\end{equation}
Under condition (A2), we have $\|\hat\bal^{(1)}_{\mathcal{A}}-\bal^{(1)}_{0,\mathcal{A}}\|_{\infty}=O_p(\sqrt{s/n})\ll l_n$. This implies that
$$\min_{j\in\mathcal{A}}|\hat\alpha^{(1)}_{j,\mathcal{A}}|>\min_{j\in\mathcal{A}}|\alpha^{(1)}_{0j,\mathcal{A}}|-l_n=l_n.$$
By the concavity of $p(\cdot)$ and condition (A2), we obtain that.
$$\|\sqrt n\lambda_n\bar{\rho}(\hat\bal^{(1)}_{\mathcal{A}})\|_2\leq (ns)^{1/2}p'_{\lambda_n}(l_n)=o(1).$$

Thus we obtain
\begin{equation}\label{eqnA.16}
\dot{L}(\hat\bvarth)=\frac{1}{\sqrt n}\sum_{i=1}^n (Y_i-\hat\eta_{i1}-\hat\bbeta^\top  \bZ_i)
\left(
  \begin{array}{ccc}
    \bZ_i+\dfrac{\partial \hat\eta_{i1}}{\partial\bbeta}\\
   \dfrac{\partial \hat\eta_{i1}}{\partial\bal^{(1)}_{\mathcal{A}}}\\
  \end{array}
\right)=o_p(1).
\end{equation}

Next we decompose $\dot{L}(\hat\bvarth)$ as follows.
\begin{eqnarray*}
\dot{L}(\hat\bvarth)&=&\frac{1}{\sqrt n}\sum_{i=1}^n [\epsilon_i+(\eta_{i0}-\hat\eta_{i0})-(\hat\eta_{i1}-\hat\eta_{i0})-(\hat\bbeta-\bbeta_0)^\top  \bZ_i]
[L_i+\hat L_{i}-L_i]\nonumber\\
&=&\frac{1}{\sqrt n}\sum_{i=1}^n \epsilon_i L_i
+\frac{1}{\sqrt n}\sum_{i=1}^n(\eta_{i0}-\hat\eta_{i0})L_i\\
&& -\frac{1}{\sqrt n}\sum_{i=1}^n[(\hat\eta_{i1}-\hat\eta_{i0})+(\hat\bbeta-\bbeta_0)^\top  \bZ_i]L_i
 +\frac{1}{\sqrt n}\sum_{i=1}^n \epsilon_i(\hat L_{i}-L_i)\\
 &&+\frac{1}{\sqrt n}\sum_{i=1}^n [(\eta_{i0}-\hat\eta_{i1})+(\bbeta_0-\hat\bbeta)^\top  \bZ_i](\hat L_{i}-L_i)\nonumber\\
&=:&F_{n1}+F_{n2}-F_{n3}+F_{n4}+F_{n5}.
\end{eqnarray*}
For the term $F_{n3}$, we have
\begin{equation}\label{eqnA.18}
F_{n3}=\frac{1}{n}\sum_{i=1}^nL_i(\widetilde L_i-L_i)^\top \sqrt n(\hat\bvarth-\bvarth_0)
+\frac{1}{n}\sum_{i=1}^nL_i L_i^\top \sqrt n(\hat\bvarth-\bvarth_0)=:F_{n31}+F_{n32}.
\end{equation}

In the following, we will show that $F_{n2}, F_{n31}, F_{n4}, F_{n5}$ are all $o_p(1)$.

Recall that $D_{n23}=(\bvarth-\bvarth_0)^\top F_{n2}/\sqrt n$. Then from the argument for the term $D_{n23}$ in the proof of step 1, we know that
\begin{eqnarray*}
F_{n2}=O_p(\sqrt s(h^2+\sqrt{\frac{\log n}{nh}}))=o_p(1),
\end{eqnarray*}
under conditions $sh^4\rightarrow 0$ and $nh/(s\log n)\rightarrow \infty$  satisfied by (A6b).

While for the term $F_{n31}$, we have
\begin{eqnarray*}
F_{n31}=O_p(s(h^2+\sqrt{\frac{\log n}{nh^3}})\sqrt s)=o_p(1),
\end{eqnarray*}
under conditions that $s^3h^4\rightarrow 0$ and $nh^3/(s^3\log n)\rightarrow \infty$ satisfied by (A6b).

From the argument for the term $D_{n22}$ in the proof of step 1,
we know that $F_{n4}$ is of the following order
\begin{eqnarray*}
F_{n4}=O_p(\sqrt s(h^2+\sqrt{\frac{\log n}{nh^3}}))=o_p(1),
\end{eqnarray*}
under conditions that $sh^4\rightarrow 0$ and $nh^3/(s\log n)\rightarrow \infty$ satisfied by (A6b).

It follows from Lemma~\ref{Lemma1} that
\begin{eqnarray*}
F_{n5}&=&O_p(\sqrt n(\sqrt{\frac{s}{n}}+h^2+\sqrt{\frac{\log n}{nh}})\sqrt{s}(h^2+\sqrt{\frac{\log n}{nh^3}}))\\
&=&O_p(sh^2+\sqrt{\frac{s^2\log n}{nh^3}}+\sqrt{ns}h^4+\sqrt{sh\log n}+\sqrt{\frac{s(\log n)^2}{nh^4}})=o_p(1),
\end{eqnarray*}
under conditions that $ s^3h^4\rightarrow 0, nh^4/(s(\log n)^2)\rightarrow\infty, nsh^8\rightarrow 0$, and $sh\log n\rightarrow 0$ satisfied by (A6b).

Thus we obtain that
\begin{eqnarray*}
o_p(1)=\dot{L}(\hat\bvarth)&=&\frac{1}{\sqrt n}\sum_{i=1}^n \epsilon_i L_i+\frac{1}{n}\sum_{i=1}^nL_i L_i^\top \sqrt n(\hat\bvarth-\bvarth_0).
\end{eqnarray*}
Recall that
$$\|\frac{1}{n}\sum_{i=1}^nL_i L_i^\top -\Sigma^{\star}\|_2=O_p(\frac{s}{\sqrt n}).$$
Thus it follows that
\begin{eqnarray*}
o_p(1)=\dot{L}(\hat\bvarth)&=&\frac{1}{\sqrt n}\sum_{i=1}^n \epsilon_i L_i+\Sigma^{\star}\sqrt n(\hat\bvarth-\bvarth_0),
\end{eqnarray*}
under condition that $s=o(n^{1/3})$.

As a result, we obtain that
\begin{equation}\label{eqnA.19}
\sqrt n(\hat\bvarth-\bvarth_0)=\Sigma^{\star -1}\frac{1}{\sqrt n}\sum_{i=1}^n \epsilon_i\left(
  \begin{array}{ccc}
    \widetilde \bZ^{\star}_i\\
   \eta'_{i0}J^\top _{0,\mathcal{A}}\widetilde \bX_i^{\star}\\
  \end{array}
\right)+o_p(1)=:\Sigma^{\star -1}\dot{L}^{\star}(\bvarth_0)+o_p(1).
\end{equation}


\vspace{5mm} {\noindent\bf{Proof of Theorem 2}:}

Similar to the arguments in the proof
of Theorem 1, we can show that $\widetilde\bal^{(1)}_{\mathcal{A}^c}=0$ with
probability 1, and further
$\|\widetilde\bal^{(1)}_{\mathcal{A}}-\bal^{(1)}_{0,\mathcal{A}}\|_2=O_p(\sqrt{s/n})$.
For $\widetilde\bvarth$, we have
\begin{equation}\label{eqnA.20}
\dot{L}(\widetilde\bvarth)=\frac{1}{\sqrt n}\sum_{i=1}^n (Y_i-\widetilde\eta_{i1}-\widetilde\bbeta^\top  \bZ_i)
\left(
  \begin{array}{ccc}
    \bZ_i+\dfrac{\partial \widetilde\eta_{i1}}{\partial\bbeta}\\
   \dfrac{\partial \widetilde\eta_{i1}}{\partial\bal^{(1)}_{\mathcal{A}}}\\
  \end{array}
\right)=\left(
  \begin{array}{ccc}
    \bm{v}\\
   \sqrt{n}\lambda_n\bar{\rho}(\widetilde\bal^{(1)}_{\mathcal{A}})\\
  \end{array}
\right).
\end{equation}
Similar to the argument for $\hat\bvarth$, we have
\begin{eqnarray*}
\sqrt n(\widetilde\bvarth-\bvarth_0)&=&\Sigma^{\star -1}
\dot{L}^{\star}(\bvarth_0)-\Sigma^{\star -1}\left(
  \begin{array}{ccc}
    I_{q\times q}\\
   0_{s\times q} \\
  \end{array}
\right)\bm{v}+o_p(1).
\end{eqnarray*}
Recall that $\widetilde\bbeta-\bbeta_0=0-\bm{\delta}_n=-\bm{\delta}_n$. Then we
have
\begin{eqnarray*}
-\sqrt n \bm{\delta}_n&=&\sqrt n(\widetilde\bbeta-\bbeta_0)=(I_q, 0_{q\times s})\sqrt n(\widetilde\bvarth-\bvarth_0)\\
&=&(I_q, 0_{q\times s})\Sigma^{\star -1}\dot{L}^{\star}(\bvarth_0)-(I_q, 0_{q\times s})\Sigma^{\star -1}\left(
  \begin{array}{ccc}
    I_{q\times q}\\
   0_{s\times q} \\
  \end{array}
\right)\bm{v}+o_p(1).
\end{eqnarray*}

Let $$\Phi=(I_q, 0_{q\times s})\Sigma^{\star -1}\left(
  \begin{array}{ccc}
    I_{q\times q}\\
   0_{s\times q} \\
  \end{array}
\right).$$
Under condition (A1), we have $\lambda_{\max}(\Sigma^{\star })=O(1)$.
This implies that $\lambda_{\min}(\Sigma^{\star -1})>0$, and then
$\lambda_{\min}(\Phi)>0$. Finally we get $\lambda_{\max}(\Phi^{-1})=O(1)$.

Then we obtain that
$$\bm{v}=\Phi^{-1}(I_q, 0_{q\times s})\Sigma^{\star -1}\dot{L}^{\star}(\bvarth_0)+\sqrt n \Phi^{-1}\bm{\delta}_n+o_p(1).$$

Consequently, it follows that
\begin{eqnarray}\label{eqnA.21}
\sqrt n(\widetilde\bvarth-\bvarth_0)&=&\Sigma^{\star -1}\dot{L}^{\star}(\bvarth_0)-\Sigma^{\star -1}\left(
  \begin{array}{ccc}
    I_{q\times q}\\
   0_{s\times q} \\
  \end{array}
\right)\Phi^{-1}(I_q, 0_{q\times s})\Sigma^{\star -1}\dot{L}^{\star}(\bvarth_0)\nonumber\\
&&-\Sigma^{\star -1}\left(
  \begin{array}{ccc}
    I_{q\times q}\\
   0_{s\times q} \\
  \end{array}
\right)\sqrt n \Phi^{-1} \bm{\delta}_n+o_p(1).
\end{eqnarray}
Or equivalently
\begin{eqnarray}\label{eqnA.22}
\sqrt n(\widetilde\bvarth-\bvarth_0)&=&\Sigma^{\star -1/2}(I-P_n)\Sigma^{\star -1/2}\dot{L}^{\star}(\bvarth_0)\nonumber\\
&&-\Sigma^{\star -1}\left(
  \begin{array}{ccc}
    I_{q\times q}\\
   0_{s\times q} \\
  \end{array}
\right)\sqrt n \Phi^{-1} \bm{\delta}_n+o_p(1).
\end{eqnarray}
Here $$P_n=\Sigma^{\star -1/2}\left(
  \begin{array}{ccc}
    I_{q\times q}\\
   0_{s\times q} \\
  \end{array}
\right)\Phi^{-1}(I_q, 0_{q\times s})\Sigma^{\star -1/2}.$$
It is easy to see that $P_n$ is an idempotent matrix with rank $q$.

From the asymptotic expansions of $\hat\bvarth$ and $\widetilde\bvarth$ in
equations (\ref{eqnA.19}) and (\ref{eqnA.22}), we have
\begin{equation}\label{eqnA.23}
\sqrt n(\hat\bvarth-\widetilde\bvarth)=\Sigma^{\star -1/2}P_n \Sigma^{\star -1/2}\dot{L}^{\star}(\bvarth_0)+\Sigma^{-1}\left(
  \begin{array}{ccc}
    I_{q\times q}\\
   0_{s\times q} \\
  \end{array}
\right)\sqrt n \Phi^{-1} \bm{\delta}_n+o_p(1).
\end{equation}

Recall that
\begin{eqnarray}\label{eqnA.24}
\dot{L}^{\star}(\bvarth_0)=\frac{1}{\sqrt n}\sum_{i=1}^n \epsilon_i\left(
  \begin{array}{ccc}
    \widetilde \bZ^{\star}_i\\
   \eta'_{i0}J^\top _{0,\mathcal{A}}\widetilde \bX_i^{\star}\\
  \end{array}
\right).
\end{eqnarray}
Then we can obtain
\begin{eqnarray*}
&&\frac{1}{\sigma^2}E\|P_n \Sigma^{\star -1/2}\dot{L}^{\star}(\bvarth_0)\|^2_2
=tr(P_n\Sigma^{\star -1/2}\Sigma^{\star}\Sigma^{\star -1/2}P_n)\\
&=&tr(P_n)=\mbox{rank}(P_n)=q.
\end{eqnarray*}
It follows that
\begin{eqnarray*}
&&E\|\Sigma^{\star -1/2}P_n \Sigma^{\star -1/2}\dot{L}^{\star}(\bvarth_0)\|^2_2
\leq \|\Sigma^{\star -1/2}\|^2_2E\|P_n \Sigma^{\star -1/2}\dot{L}^{\star}(\bvarth_0)\|^2_2=O(1).
\end{eqnarray*}
Consequently, under condition (A7), we have
\begin{eqnarray}\label{eqnA.25}
\sqrt n(\hat\bvarth-\widetilde\bvarth)=O_p(1).
\end{eqnarray}
Now we are ready to investigate the asymptotic distribution of the F-type test $T_n$. Let $D(\bvarth^*)=n^{-1}\sum_{i=1}^n L_i^*L_i^{*\top}$.
Under the event $\hat\bal^{(1)}_{\mathcal{A}^c}=\widetilde\bal^{(1)}_{\mathcal{A}^c}=0$ and
recalling  (\ref{eqnA.15}), we obtain that
\begin{eqnarray}\label{eqnA.26}
&&\frac{1}{2}(RSS_0-RSS_1)\nonumber\\
&=&-(\widetilde\bal^{(1)}_{\mathcal{A}}-\hat\bal^{(1)}_{\mathcal{A}})^\top n\lambda_n\bar{\rho}(\hat\bal^{(1)}_{\mathcal{A}})
+\frac{1}{2}(\widetilde\bvarth-\hat\bvarth)^\top nD(\bvarth^*)(\widetilde\bvarth-\hat\bvarth)
\nonumber\\
&=&\frac{1}{2}\sqrt n(\hat\bvarth-\widetilde\bvarth)^\top \Sigma^{\star}\sqrt n(\hat\bvarth-\widetilde\bvarth)+o_p(1)\nonumber\\
&=&\frac{1}{2}\|P_n \Sigma^{\star -1/2}\dot{L}^{\star}(\bvarth_0)+\Sigma^{\star -1/2}\left(\begin{array}{ccc}
    I_q\\
   0_{s\times q}\\
  \end{array}
\right)\sqrt n \Phi^{-1} \bm{\delta}_n\|^2_2+o_p(1).
\end{eqnarray}
The second equation follows from (\ref{eqnA.25}) and $n\lambda_n\bar{\rho}(\hat\bal^{(1)}_{\mathcal{A}})=o_p(n^{1/2})$ based on condition (A2).
The last equation holds due to equation (\ref{eqnA.22}).
Recall that
$$P_n=\Sigma^{\star -1/2}\left(
  \begin{array}{ccc}
    I_{q\times q}\\
   0_{s\times q} \\
  \end{array}
\right)\Phi^{-1}(I_q, 0_{q\times s})\Sigma^{\star -1/2}.$$
Further denote that
$$\bm{\omega}_n=(I_q, 0_{q\times s})\Sigma^{\star -1}\dot{L}^{\star }(\bvarth_0).$$
It follows that
\begin{eqnarray}\label{eqnA.27}
&&\|P_n \Sigma^{\star -1/2}\dot{L}^{\star}(\bvarth_0)+\Sigma^{\star -1/2}\left(\begin{array}{ccc}
    I_q\\
   0_{s\times q}\\
  \end{array}
\right)\sqrt n \Phi^{-1} \bm{\delta}_n\|^2_2\nonumber\\
&=&\|\Sigma^{\star -1/2}\left(\begin{array}{ccc}
    I_q\\
   0_{s\times q}\\
  \end{array}
\right)\Phi^{-1}\bm{\omega}_n+\Sigma^{\star -1/2}\left(\begin{array}{ccc}
    I_q\\
   0_{s\times q}\\
  \end{array}
\right)\sqrt n \Phi^{-1} \bm{\delta}_n\|^2_2\nonumber\\
&=&\|\Phi^{-1/2}\bm{\omega}_n+\sqrt n \Phi^{-1/2} \bm{\delta}_n\|^2_2.
\end{eqnarray}
Thus we obtain that
\begin{eqnarray}\label{eqnA.28}
RSS_0-RSS_1&=&\|\Phi^{-1/2}\bm{\omega_n}+\sqrt n \Phi^{-1/2} \bm{\delta}_n\|^2_2+o_p(1).
\end{eqnarray}
It is easy to know that $\Phi^{-1/2}\bm{\omega}_n\rightarrow N(0,\sigma^2I_q)$.

In the following, we aim to show that $RSS_1/(n-q)$ is a consistent estimator
of $\sigma^2$. In fact, we have
\begin{eqnarray*}
\frac{RSS_1}{n-q}=\frac{1}{n-q}\sum_{i=1}^n[Y_i-\hat\eta_{i1}-\hat\bbeta^\top  \bZ_i]^2.
\end{eqnarray*}
Due to the consistencies of the related estimators, it is clear that
\begin{eqnarray*}
Y_i-\hat\eta_{i1}-\hat\bbeta^\top  \bZ_i=\epsilon_i+o_p(1).
\end{eqnarray*}
Thus we obtain that
\begin{eqnarray*}
\frac{RSS_1}{n-q}=\frac{1}{n-q}\sum_{i=1}^n\epsilon_i^2+o_p(1)\rightarrow\sigma^2.
\end{eqnarray*}
As a result, we have
\begin{eqnarray*}
T_n=\frac{RSS_0-RSS_1}{RSS_1/(n-q)}&=&{\|\Phi^{-1/2}\bm{\omega}_n/\sigma+\sqrt n \Phi^{-1/2} \bm{\delta}_n/\sigma\|^2_2}+o_p(1).
\end{eqnarray*}
Thus $$T_n\rightarrow \chi^2_{q}(n\bm{\delta}_n^\top \Phi^{-1}\bm{\delta}_n/\sigma^2).$$

\vspace{3mm}

\vspace{5mm} {\noindent\bf{Proof of Theorem 3}:}

Under the null hypothesis,
\begin{eqnarray*}
\hat\epsilon_{0i}=\epsilon_{0i}+[\bbeta^\top _0 \bZ_i-\hat\bbeta^\top \bZ_i+g(\bal^\top _0 \bX_i,\zeta_0)-g(\hat \bal^\top  \bX_i,\hat\zeta)]
=:\epsilon_{0i}+\Delta_i.
\end{eqnarray*} Further denote $\bal_{ij}=\bal^\top _0(\bX_i-\bX_j)$. Thus we have
\begin{eqnarray}\label{eqnA.29}
S_n&=&\frac{1}{n(n-1)}\sum_{i=1}^n\sum_{j\neq i}^n[\epsilon_{0i}\epsilon_{0j}+(\epsilon_{0i}\Delta_j+\epsilon_{0j}\Delta_i)
+\Delta_i\Delta_j]\frac{1}{b}G(\frac{\hat\bal_{ij}}{b})\nonumber\\
&=&S_{n1}+S_{n2}+S_{n3}.
\end{eqnarray}
For the first term $S_{n1}$, we have
\begin{eqnarray*}
S_{n1}&=&\frac{1}{n(n-1)}\sum_{i=1}^n\sum_{j\neq i}^n\epsilon_{0i}\epsilon_{0j}
\frac{1}{b}G(\frac{\bal_{ij}}{b})\\
&&+\frac{1}{n(n-1)}\sum_{i=1}^n\sum_{j\neq i}^n\epsilon_{0i}\epsilon_{0j}\frac{1}{b}[G(\frac{\hat\bal_{ij}}{b})-G(\frac{\bal_{ij}}{b})]\\
&=&:S_{n11}+S_{n12}.
\end{eqnarray*}
Note that $$E[\epsilon_{0i}\epsilon_{0j}G(\frac{\bal_{ij}}{b})|\epsilon_{0i},
\bal^\top _0 \bX_i]
=E[E(\epsilon_{0i}\epsilon_{0j}G(\frac{\bal_{ij}}{b})|\epsilon_{0i},
\bal^\top _0 \bX_i,\bal^\top _0 \bX_j)|\epsilon_{0i}, \bal^\top _0 \bX_i]=0.$$
Thus $S_{n11}$ is a degenerate U-statistic. From \cite{zheng1996consistent}, we get
\begin{eqnarray}\label{eqnA.30}
nb^{1/2}S_{n11}\rightarrow N(0,\sigma^2_S).
\end{eqnarray}
Here
$$\sigma^2_S=2\int G^2(t)dt\cdot\int \sigma^4 f^2(\bal^\top _0 \bX)d\bal^\top _0 \bX. $$

Next, we aim to show that $S_{n12}, S_{n2}, S_{n3}$ are all of
order $o_p((nb^{1/2})^{-1})$.

Denote
$$S^*_{n12}=\frac{1}{n(n-1)}\sum_{i=1}^n\sum_{j\neq i}^n\epsilon_{0i}
\epsilon_{0j}\frac{1}{b}G'(\frac{\bal_{ij}}{b})(\bX_i-\bX_j)^\top \frac{\hat
\bal-\bal}{b}.$$ Clearly, we have
\begin{eqnarray*}
S_{n12}&=&S^*_{n12}+o_p(S^*_{n12}).
\end{eqnarray*}
Since $G(\cdot)$ is a symmetric function, similar to $S_{n11}$, the following term
\begin{eqnarray*}
\widetilde S_n=\frac{1}{n(n-1)}\sum_{i=1}^n\sum_{j\neq i}^n\epsilon_{0i}\epsilon_{0j}\frac{1}{b}G'(\frac{\bal_{ij}}{b})(X_{ik}-X_{jk})
\end{eqnarray*}
is also a degenerate U-statistic. To determine its order, we can compute its
second order moment as follows
\begin{eqnarray*}
E(\widetilde S^2_n)&=&\frac{1}{n^2(n-1)^2}\sum_{i=1}^n\sum_{j\neq i}^n
\sum_{i'=1}^n\sum_{j'\neq i'}^nE\{\epsilon_{0i}\epsilon_{0j}\epsilon_{0i'}\epsilon_{0j'}
\frac{1}{b^2}G'(\frac{\bal_{ij}}{b})(X_{ik}-X_{jk})\\
&& \times G'(\frac{\bal_{i'j'}}{b})(X_{i'k}-X_{j'k})\}.
\end{eqnarray*}
Since $E(\epsilon_{0i}|\bX_i)=0$, we only need to consider the
terms with $i=i'\neq j=j'$ or $i=j'\neq j=i'$. Then, it follows that
\begin{eqnarray*}
E(\widetilde S^2_n)&=&\frac{2}{n^2(n-1)^2}\sum_{i=1}^n\sum_{j\neq i}^n
E[\epsilon^2_{0i}\epsilon^2_{0j}\frac{1}{b^2}G'^2(\frac{\bal_{ij}}{b})(X_{ik}-X_{jk})^2]\\
&=&\frac{2\sigma^4}{n^2(n-1)^2}\sum_{i=1}^n\sum_{j\neq i}^n
E\Big[E((X_{ik}-X_{jk})^2|\bal_{ij})\frac{1}{b^2}G'^2(\frac{\bal_{ij}}{b})\Big]=O(\frac{1}{n^2b}).
\end{eqnarray*}
Consequently, we have that $\widetilde S_n=O_p((nb^{1/2})^{-1}).$
Under the event $\hat\bal_{\mathcal{A}^c}=0$ with probability tending to 1, we obtain that
\begin{eqnarray}\label{eqnA.31}
S_{n12}=O_p(\sqrt s\frac{1}{nb^{1/2}}\sqrt{\frac{s}{n}}\frac{1}{b})=o_p(\frac{1}{nb^{1/2}}),
\end{eqnarray}
under condition that $nb^2/s^2\rightarrow \infty$.
Denote $\gamma=(\bbeta^\top , \bal^\top _{\mathcal{A}}, \zeta)^\top $ and $\bM_i=\left(         \begin{array}{c}
                                                \bZ_i \\
                                                g_{10}(\bal^\top _0\bX_i,\zeta_0)\bX_{i,\mathcal{A}}\\
                                                g_{01}(\bal^\top _0\bX_i,\zeta_0) \\
                                            \end{array}
                                            \right)$.
Further let
$$S_{n2}^*=2\frac{(\gamma_0-\hat\gamma)^\top }{n(n-1)}\sum_{i=1}^n\sum_{j\neq i}^n\epsilon_{0i}\bM_j\frac{1}{b}G(\frac{\hat\bal_{ij}}{b})$$
Clearly, we have
\begin{eqnarray*}
S_{n2}&=&S^*_{n2}+o_p(S^*_{n12}).
\end{eqnarray*}

Similar to the argument for $S_{n1}$ and from Lemma 2 in \cite{guo2016model} and Lemma 2, we can derive that
\begin{eqnarray}\label{eqnA.32}
S_{n2}=O_p(\sqrt{\frac{s}{n}}\sqrt{\frac{s}{n}})=o_p(\frac{1}{nb^{1/2}}),
\end{eqnarray}
under condition that $sb^{1/2}\rightarrow 0$.

Further let
\begin{eqnarray*}
S^*_{n3}&=&(\gamma_0-\hat\gamma)^\top \frac{1}{n(n-1)}\sum_{i=1}^n\sum_{j\neq i}^n\bM_i\bM^\top _j\frac{1}{b}G(\frac{\hat\bal_{ij}}{b})(\gamma_0-\hat\gamma)
\end{eqnarray*}
Under assumption that $\lambda_{\max}(E[\bM\bM^\top])<\infty$ and based on Lemma 2, we can also obtain that
\begin{eqnarray}\label{eqnA.33}
S_{n3}=O_p(\sqrt{\frac{s}{n}}\sqrt{\frac{s}{n}})=o_p(\frac{1}{nb^{1/2}}).
\end{eqnarray}
In sum, under the null hypothesis with conditions that $nb^2/s^2\rightarrow \infty$ and $sb^{1/2}\rightarrow 0$, we obtain that
$$nb^{1/2}S_n\rightarrow N(0,\sigma^2_S). $$
Since $\sigma^2_S$ is actually unknown, an estimate is defined as
$$\hat\sigma^2_S=\frac{2}{n(n-1)}\sum_{i=1}^n\sum_{j\neq i}^n
\frac{1}{b}G^2(\frac{\hat\bal^\top (\bX_i-\bX_j)}{b})\hat\epsilon^2_{0i}\hat\epsilon^2_{0j}.$$
The proof follows from the U-statistic theory and the consistencies of
parametric estimators, and thus the details are omitted here.

\vspace{3mm}
The following two Lemmas are used in the proof of the main Theorems.
We first present the following lemma,
\begin{lemma}
Under conditions (A4) and (A5), for any $\bvarth$ which satisfies $\|\bvarth-\bvarth_0\|_2=O(\sqrt{s/n})$,
we have
\begin{eqnarray*}
E[\hat \eta_{i1}-\eta_{i0}]^2=O(\frac{s}{n}+h^4+\frac{1}{nh});\,\,\,E\|\hat L_i-L_i\|^2_2=O(s(h^4+\frac{1}{{nh^3}})).
\end{eqnarray*}
\end{lemma}

{\emph{Proof}:} Since the proof for the second statement is more complicated, we only focus on the second result. The first result can be similarly demonstrated and thus omitted here.
Recall that
\begin{equation}\label{eqnL.2}
\hat\eta_{i1}=\hat\eta(\bal^\top \bX_i,\btheta)=\frac{T_{20}(\bX_i,\btheta)T_{01}(\bX_i,\btheta)-T_{10}(\bX_i,\btheta)T_{11}(\bX_i,\btheta)}
{T_{00}(\bX_i,\btheta)T_{20}(\bX_i,\btheta)-T^2_{10}(\bX_i,\btheta)},
\end{equation}
where
\begin{eqnarray*}
T_{l_1,l_2}(\bX_i,\btheta)=\sum_{j\neq i}^n K_h(\Gamma_{j1}-\Gamma_{i1})(\Gamma_{j1}-\Gamma_{i1})^{l_1}(Y_j-\bbeta^\top \bZ_j)^{l_2},
\end{eqnarray*}
for $l_1=0,1,2$ and $l_2=0,1$.

Define
\begin{eqnarray*}
&&G_{n1}(\bX_i,\btheta)=\frac{1}{nh^2}T_{20}(\bX_i,\btheta)\frac{1}{n}T_{01}(\bX_i,\btheta)-\frac{1}{nh}T_{10}(\bX_i,\btheta)\frac{1}{nh}T_{11}(\bX_i,\btheta);\\
&&G_{n2}(\bX_i,\btheta)=\frac{1}{nh^2}T_{20}(\bX_i,\btheta)\frac{1}{n}T_{00}(\bX_i,\btheta)-\frac{1}{n^2h^2}T^2_{10}(\bX_i,\btheta).
\end{eqnarray*}
Clearly,
\begin{eqnarray*}
\frac{\partial \hat\eta_{i1}}{\partial\bal^{(1)}_{\mathcal{A}}}
=\frac{\partial G_{n1}(\bX_i,\btheta)/\partial\bal^{(1)}_{\mathcal{A}}}{G_{n2}(\bX_i,\btheta)}
-\frac{G_{n1}(\bX_i,\btheta)\partial G_{n2}(\bX_i,\btheta)/\partial\bal^{(1)}_{\mathcal{A}}}{G^2_{n2}(\bX_i,\btheta)}.
\end{eqnarray*}
Further
\begin{align}\label{eqnL.3}
&&\frac{\partial G_{n1}(\bX_i,\btheta)}{\partial\bal^{(1)}_{\mathcal{A}}}
=\frac{1}{nh^2}\frac{\partial T_{20}(\bX_i,\btheta)}{\partial\bal^{(1)}_{\mathcal{A}}}
\frac{1}{n}T_{01}(\bX_i,\btheta)+\frac{1}{nh^2}T_{20}(\bX_i,\btheta)\frac{1}{n}\frac{\partial T_{01}(\bX_i,\btheta)}{\partial\bal^{(1)}_{\mathcal{A}}}\nonumber\\
&&-\frac{1}{nh}\frac{\partial T_{10}(\bX_i,\btheta)}{\partial\bal^{(1)}_{\mathcal{A}}}
\frac{1}{nh}T_{11}(\bX_i,\btheta)-\frac{1}{nh}T_{10}(\bX_i,\btheta)\frac{1}{nh}\frac{ \partial T_{11}(\bX_i,\btheta)}{\partial\bal^{(1)}_{\mathcal{A}}}.
\end{align}

In the following, we only deal with the first term of $\partial G_{n1}(\bX_i,\btheta)/\partial\bal^{(1)}_{\mathcal{A}}$.

Recall that
\begin{eqnarray*}
\frac{1}{nh^2}T_{20}(\bX_i,\btheta)=\frac{1}{n}\sum_{j\neq i}^n \frac{1}{h}K(\frac{\Gamma_{j1}-\Gamma_{i1}}{h})[\frac{\Gamma_{j1}-\Gamma_{i1}}{h}]^{2}
=f(\Gamma_{i1})\mu_{K2}+O_p(h^2+\frac{1}{\sqrt{nh}}).
\end{eqnarray*}
Here $\mu_{K2}=\int K(t)t^2 dt$.
Notice that
\begin{eqnarray*}
&&\frac{1}{nh^2}\frac{\partial T_{20}(\bX_i,\btheta)}{\partial\bal^{(1)}_{\mathcal{A}}}
=\frac{1}{n}\sum_{j\neq i}^n \frac{1}{h}K'(\frac{\Gamma_{j1}-\Gamma_{i1}}{h})[\frac{\Gamma_{j1}-\Gamma_{i1}}{h}]^{2}
J_{\mathcal{A}}^\top  \frac{\bX_{j,\A^*}-\bX_{i,\A^*}}{h}\\
&&+\frac{2}{n}\sum_{j\neq i}^n \frac{1}{h}K(\frac{\Gamma_{j1}-\Gamma_{i1}}{h})\frac{\Gamma_{j1}-\Gamma_{i1}}{h}J_{\mathcal{A}}^\top  \frac{\bX_{j,\A^*}-\bX_{i,\A^*}}{h}=:A_1+A_2.
\end{eqnarray*}
For the term $A_1$, we have:
\begin{eqnarray*}
A_1&=&\frac{1}{nh}\sum_{j\neq i}^n \frac{1}{h}K'(\frac{\Gamma_{j1}-\Gamma_{i1}}{h})[\frac{\Gamma_{j1}-\Gamma_{i1}}{h}]^{2}
J_{\mathcal{A}}^\top  \bX_{j,\A^*}\\
&&-J_{\mathcal{A}}^\top \bX_{i,\A^*}\frac{1}{nh}\sum_{j\neq i}^n \frac{1}{h}K'(\frac{\Gamma_{j1}-\Gamma_{i1}}{h})[\frac{\Gamma_{j1}-\Gamma_{i1}}{h}]^{2}
=:A_{11}-A_{12}.
\end{eqnarray*}
Now we determine the expectation and variance of $A_{11}$ and $A_{12}$.
In fact, under conditions (A4) and (A5), we have:
\begin{eqnarray*}
&&E[A_{11}|\bX_i]\\
&=&\frac{J_{\mathcal{A}}^\top  }{h}\int \frac{1}{h}K'(\frac{\Gamma_{j1}-\Gamma_{i1}}{h})[\frac{\Gamma_{j1}-\Gamma_{i1}}{h}]^2
\mu_1(\Gamma_{j1})f(\Gamma_{j1})d \Gamma_{j1}\\
&=&\frac{J_{\mathcal{A}}^\top  }{h}\int K'(t)t^2
(\mu_1f)(\Gamma_{i1}+ht)dt\\
&=&\frac{J_{\mathcal{A}}^\top  }{h}\int K'(t)t^2
[(\mu_1f)(\Gamma_{i1})+(\mu_1f)'(\Gamma_{i1})ht+\\
&&\frac{(\mu_1f)''(\Gamma_{i1})}{2}h^2t^2
+\frac{(\mu_1f)''(\widetilde \Gamma_{i1})-(\mu_1f)''(\Gamma_{i1})}{2}h^2t^2]dt\\
&=&J_{0,\mathcal{A}}^\top (\mu_1f)'(\Gamma_{i0})\int K'(t)t^3dt+O(\sqrt s h^2+\sqrt{s/n}).
\end{eqnarray*}
Similarly, we get
\begin{eqnarray*}
&&E[\|A_{11}-E[A_{11}|\bX_i]\|^2_2|\bX_i]=O(\frac{s}{nh^3}).
\end{eqnarray*}
Next note that:
\begin{eqnarray*}
&&E[\frac{1}{nh}\sum_{j\neq i}^n \frac{1}{h}K'(\frac{\Gamma_{j1}-\Gamma_{i1}}{h})[\frac{\Gamma_{j1}-\Gamma_{i1}}{h}]^{2}|\bX_i]=f'(\Gamma_{i1})\int K'(t)t^3 dt+O(h^2);\\
&&Var[\frac{1}{nh}\sum_{j\neq i}^n \frac{1}{h}K'(\frac{\Gamma_{j1}-\Gamma_{i1}}{h})[\frac{\Gamma_{j1}-\Gamma_{i1}}{h}]^{2}|\bX_i]=O(\frac{1}{nh^3});
\end{eqnarray*}
In sum, we get:
\begin{equation*}
A_1=J_{0,\mathcal{A}}^\top \{[\mu_1(\Gamma_{i0})-\bX_{i,\A^*}]f'(\Gamma_{i0})+\mu'_1(\Gamma_{i0})f(\Gamma_{i0})\}\int K'(t)t^3 dt
+O_p(\sqrt s(h^2+\frac{1}{\sqrt{nh^3}})).
\end{equation*}
Similarly, we obtain that
\begin{equation*}
A_2=2J_{0,\mathcal{A}}^\top \{[\mu_1(\Gamma_{i0})-\bX_{i,\A^*}]f'(\Gamma_{i0})+\mu'_1(\Gamma_{i0})f(\Gamma_{i0})\}\int K(t)t^2 dt
+O_p(\sqrt s(h^2+\frac{1}{\sqrt{nh^3}})).
\end{equation*}
Note that $\int K'(t)t^3 dt=-3\int K(t)t^2 dt$. Consequently, we get:
\begin{equation*}
\frac{1}{nh^2}\frac{\partial T_{20}(\bX_i,\btheta)}{\partial\bal^{(1)}_{\mathcal{A}}}=-J_{0,\mathcal{A}}^\top \{[\mu_1(\Gamma_{i0})-\bX_{i,\A^*}]f'(\Gamma_{i0})+\mu'_1(\Gamma_{i0})f(\Gamma_{i0})\}\mu_{K2}
+O_p(\sqrt s(h^2+\frac{1}{\sqrt{nh^3}})).
\end{equation*}

Next we turn to consider the term $n^{-1}T_{01}(\bX_i,\btheta)$. Denote $\Delta_{\alpha}=\bal^{(1)}_{0,\mathcal{A}}-\bdelta$ and $\Delta_{\beta}=\bbeta_0-\bbeta$, which are both of order $\sqrt{s/n}$. Further note that
$$Y_j-\bbeta^\top \bZ_j=\eta_{j1}+\eta'_{j1}\bX^\top _{j,\A^*}J_{\mathcal{A}}\Delta_{\alpha}
+\frac{\eta''(\bX^\top _{j}\widetilde\bal)}{2}[\bX^\top _{j,\A^*}J_{\mathcal{A}}\Delta_{\alpha}]^2+ \bZ_j^\top \Delta_{\beta}+\epsilon_j.$$

Then we get:
\begin{eqnarray*}
&&\frac{1}{n}T_{01}(\bX_i,\btheta)=\frac{1}{n}\sum_{j\neq i}^n \frac{1}{h}K(\frac{\Gamma_{j1}-\Gamma_{i1}}{h})(Y_j-\bbeta^\top \bZ_j)\\
&=&\frac{1}{nh}\sum_{j\neq i}^n K(\frac{\Gamma_{j1}-\Gamma_{i1}}{h})
[\eta_{j1}+\eta'_{j1}\bX^\top _{j,\A^*}J_{\mathcal{A}}\Delta_{\alpha}
+\frac{\eta''(\bX^\top _{j}\widetilde\bal)}{2}[\bX^\top _{j,\A^*}J_{\mathcal{A}}\Delta_{\alpha}]^2+\bZ_j^\top \Delta_{\beta}+\epsilon_j]\\
&=&f(\Gamma_{i1})[\eta_{i1}+\eta'_{i1}\mu^\top _1(\Gamma_{i1})J_{\mathcal{A}}\Delta_{\alpha}
+\mu^\top _2(\Gamma_{i1})\Delta_{\bbeta}]
+O_p(c_n+c_n\sqrt{\frac{s^2}{n}}
+\frac{s}{n}+\frac{s^2}{n}c_n)\\
&=&f(\Gamma_{i0})\eta_{i0}+O_p(\sqrt{\frac{s}{n}}+c_n),
\end{eqnarray*}
under condition that $s=o(n^{1/2})$.

Further note that
$$O_p(\sqrt{\frac{s}{n}}+c_n)=o_p(\sqrt s(h^2+\frac{1}{\sqrt{nh^3}})).$$
Then we have:
\begin{eqnarray*}
&&\frac{1}{nh^2}\frac{\partial T_{20}(\bX_i,\btheta)}{\partial\bal^{(1)}_{\mathcal{A}}}
\frac{1}{n}T_{01}(\bX_i,\btheta)\\
&=&-J_{0,\mathcal{A}}^\top \{[\mu_1(\Gamma_{i0})-\bX_{i,\A^*}]f'(\Gamma_{i0})+\mu'_1(\Gamma_{i0})f(\Gamma_{i0})\}f(\Gamma_{i0})\eta_{i0}\mu_{K2}
+O_p(\sqrt s(h^2+\frac{1}{\sqrt{nh^3}})).
\end{eqnarray*}
Other terms of $\partial G_{n1}(\bX_i,\btheta)/\partial\bal^{(1)}_{\mathcal{A}}$ and also $\partial \hat\eta_{i1}/\partial\bal^{(1)}_{\mathcal{A}}$
can be handled similarly. After tedious calculations, we finally get:
\begin{eqnarray}\label{eqnL.4}
\frac{\partial \hat\eta_{i1}}{\partial\bal^{(1)}_{\mathcal{A}}}
=\eta'_{i0}J^\top _{0,\mathcal{A}}\widetilde \bX_i^{\star}+O_p(\sqrt s(h^2+\frac{1}{\sqrt{nh^3}})).
\end{eqnarray}

Further note that
\begin{eqnarray*}
\frac{\partial \hat\eta_{i1}}{\partial\bbeta}=\sum_{j\neq i} W_{1ij}\bZ_j=\mu_2(\Gamma_{i1})+O_p(h^2+\frac{1}{\sqrt{nh}})
=\mu_2(\Gamma_{i0})+O_p(\sqrt{\frac{s}{n}}+h^2+\frac{1}{\sqrt{nh}}).
\end{eqnarray*}

Then eventually we obtain that
$$E\|\hat L_i-L_i\|^2_2=O(s(h^4+\frac{1}{{nh^3}})).$$

\vspace{3mm}

We present the following lemma about the convergence rate of $\hat\zeta-\zeta_0$:
\begin{lemma}
Under conditions (B2) and (B2), and the assumption that $\|\hat\btheta-\btheta_0\|_2=O_p(\sqrt{s/n})$, we have:
\begin{eqnarray*}
\|\hat\zeta-\zeta_0\|_2=O_p(\sqrt{\frac{s}{n}}).
\end{eqnarray*}
\end{lemma}

{\emph{Proof:}} In fact, the proof follows from the proof for Lemma 4.2 in \cite{van2008goodness}.
From \cite{van2008goodness}, we know that the convergence rate of $\sqrt{n}(\hat\zeta-\zeta_0)$ is determined by the following term:
$$C_n=\frac{1}{\sqrt n}\sum_{i=1}^n [Y_i-\hat\bbeta^\top \bZ_i-g(\hat\bal^\top \bX_i,\zeta_0)]g_{01}(\hat\bal^\top \bX_i,\zeta_0).$$
Here $g_{01}(\cdot,\zeta_0)=\partial g(\cdot,\zeta)/\partial \zeta|_{\zeta=\zeta_0}$.

First note that under the event $\hat\bal_{\mathcal{A}^c}=0$ with probability tending to 1, we have:
\begin{eqnarray*}
&&Y_i-\hat\bbeta^\top \bZ_i-g(\hat\bal^\top \bX_i,\zeta_0)\\
&=&\epsilon_i-(\hat\bbeta-\bbeta_0)^\top \bZ_i-g_{10}(\bal^\top _0\bX_i,\zeta_0)(\hat\bal-\bal_0)^\top \bX_i
-g_{20}(\bal^{*\top}\bX_i,\zeta_0)[(\hat\bal-\bal_0)^\top \bX_i]^2/2\\
&=&\epsilon_i-(\hat{\bm{\iota}}-\bm{\iota}_0)^\top N_i-g_{20}(\bal^{*\top}\bX_i,\zeta_0)[(\hat\bal_{\mathcal{A}}-\bal_{0,\mathcal{A}})^\top \bX_{i,\mathcal{A}}]^2/2.
\end{eqnarray*}
Here $g_{k0}(\bal^\top _0\bX_i,\cdot)=\partial^k g(\bal^\top \bX_i,\cdot)/\partial^k \bal^\top \bX_i|_{\bal=\bal_0},k=1,2, \bm{\iota}=(\bbeta,\bal_{\mathcal{A}}), \bm{\iota}_0$ and $\hat{\bm{\iota}}$ are similarly defined and $\bN_i=(\bZ_i^\top , g_{10}(\bal^\top _0\bX_i,\zeta_0)\bX^\top _{i,\mathcal{A}})^\top $.

Secondly
\begin{eqnarray*}
&&g_{01}(\hat\bal^\top \bX_i,\zeta_0)=g_{01}(\bal^\top _0\bX_i,\zeta_0)+[g_{01}(\hat\bal^\top \bX_i,\zeta_0)-g_{01}(\bal^\top _0\bX_i,\zeta_0)].
\end{eqnarray*}

Thus we get
\begin{eqnarray*}
C_n&=&\frac{1}{\sqrt n}\sum_{i=1}^n\epsilon_ig_{01}(\hat\bal^\top \bX_i,\zeta_0)-(\hat{\bm{\iota}}-\bm{\iota}_0)^\top \frac{1}{\sqrt n}\sum_{i=1}^n\bN_i
g_{01}(\hat\bal^\top \bX_i,\zeta_0)\\
&&-\frac{1}{2\sqrt n}\sum_{i=1}^ng_{20}(\bal^{*\top}\bX_i,\zeta_0)[(\hat\bal-\bal_0)^\top \bX_i]^2g_{01}(\hat\bal^\top \bX_i,\zeta_0).
\end{eqnarray*}
Under the assumption that $g_{20}(\bal^\top \bX_i,\cdot)$ is bounded and $g_{01}(\bal^\top \bX_i,\zeta_0)$ satisfies Lipschitz condition of order 1 for $\bal^\top \bX_i$ in a neighborhood of $\bal^\top _0 \bX_i$, it is known that the order is determined by the second term.

We note that for any $\bm{\iota}$ which satisfies that $\|\bm{\iota}-\bm{\iota}_0\|_2=O_p(\sqrt{s/n})$,
\begin{eqnarray*}
E[({\bm{\iota}}-\bm{\iota}_0)^\top \bN_ig_{01}(\bal^\top _0\bX_i,\zeta_0)]^2=({\bm{\iota}}-\bm{\iota}_0)^\top E[g^2_{01}(\bal^\top _0\bX_i,\zeta_0)\bN_i\bN^\top _i]
({\bm{\iota}}-\bm{\iota}_0)=O_(\frac{s}{n}).
\end{eqnarray*}
The last equation holds under condition that $\lambda_{\max}(E[g^2_{01}(\bal^\top _0\bX_i,\zeta_0)\bN_i\bN^\top _i])<\infty$. Thus the results follow.



\bibliographystyle{asa} 
\bibliography{bibliography}       


\end{document}